\documentclass[sigconf, 10pt]{acmart}
\usepackage{etoolbox}
\makeatother
\renewcommand\footnotetextcopyrightpermission[1]{} 

\usepackage{booktabs,xcolor,hyperref,enumerate}
\usepackage[T1]{fontenc}
\usepackage[utf8]{inputenc}
\usepackage{bm,courier,xspace,fancyhdr,fancyref,wrapfig}
\usepackage{amsmath,amsfonts,tabularx,tablefootnote,wrapfig,hyphenat}
\usepackage[labelformat=simple,skip=0pt]{subcaption}
\usepackage{xfrac,algorithm,algpseudocode}
\usepackage{paralist,mathtools,upgreek}
\usepackage{float,enumitem}
\usepackage[capitalize]{cleveref}
\usepackage{gensymb}
\usepackage[export]{adjustbox}
\usepackage{threeparttable}
\usepackage{multirow}
\usepackage[table]{xcolor} 
\usepackage{xurl}

\DeclareCaptionLabelSeparator{emdash}{---~}
\microtypecontext{spacing=nonfrench}

\usepackage[toc]{appendix}

\newcommand{\parahead}[1]{\noindent{}{\bfseries #1}}

\let\oldReturn\Return
\renewcommand{\Return}{\State\oldReturn}

\newcommand{\shortname}{Wall-Street}
\newcommand{\systemname}{\shortname{}}
\newcommand{\shortnames}{Wall-Street's}
\newcommand{\systemnames}{\shortnames{}}

\setitemize{itemsep=2pt,topsep=2pt,parsep=1pt,partopsep=0pt,leftmargin=1em}
\setenumerate{itemsep=3pt,topsep=4pt,parsep=1pt,partopsep=0pt,leftmargin=1em}

\crefformat{section}{\S#2#1#3}
\Crefformat{section}{Section~#2#1#3}
\crefname{figure}{Fig.}{Figs.}
\Crefname{figure}{Figure}{Figures}
\crefformat{equation}{Eq.~(#2#1#3)}
\crefmultiformat{equation}{Eqs.~(#2#1#3)}{ and~(#2#1#3)}{, (#2#1#3)}{ and~(#2#1#3)}

\begin{document}

\title{Wall-Street: An Intelligent Vehicular Surface for Reliable mmWave Handover}
\author{{Kun Woo Cho$^{1,2}$, Prasanthi Maddala$^3$, Ivan Seskar$^3$, Kyle Jamieson$^1$}}
\affiliation{Princeton University$^1$, Rice University$^2$, Rutgers University$^3$}


\begin{abstract}

mmWave networks promise high bandwidth but face significant challenges in maintaining reliable connections for users moving at high speed. 
Frequent handovers, complex beam alignment, and signal blockage from car bodies lead to service interruptions and degraded performance. 
We present \textbf{\shortname{},} a vehicle-mounted smart surface that enhances mmWave connectivity for in-vehicle users.  
\shortname{} improves mobility management by 
(1) steering outdoor mmWave signals into the vehicle for shared coverage and providing a single, collective handover for all users; 
(2) performing neighbor-cell search without interrupting data transfer, ensuring seamless handovers; and 
(3) connecting users to a new cell before disconnecting from the old cell for reliable cell transitions.
We implemented and integrated \shortname{} into the COSMOS testbed. 
We collected PHY traces with multiple base station nodes and in-vehicle user nodes with a surface-mounted vehicle, driving on a nearby road. 
Our trace-driven ns-3 simulation demonstrates a throughput improvement of up to 78\% and a latency reduction of up to 34\% over the standard Standalone handover scheme.\footnote{This work was done when Kun Woo Cho was a student at Princeton University.}



\end{abstract}

\maketitle
\pagestyle{plain}

\section{Introduction}
\label{s:intro}

5G wireless networks leverages millimeter wave (mmWave) spectrum in the FR2 band to deliver multi-gigabit, low-latency connectivity \cite{3gpp_release_2022}. 
However, due to its short wavelengths and limited diffraction capability, 
mmWave spectrum suffers from high path loss and extreme vulnerability to blockage, posing significant challenges in maintaining reliable connections, particularly in high mobility scenarios \cite{rappaport_small-scale_2017}.

To overcome poor signal propagation, mmWave networks require dense deployments of small cells, which in turn leads to handovers (HOs) that are far more frequent than in low-band networks.
Indeed, mmWave 5G NR experiences a HO significantly more often than LTE or low band 5G NR \cite{hassan_vivisecting_2022, narayanan_variegated_2021}. 
The mmWave HO process adds a layer of complexity on top of low-band procedures \cite{deng_mobility_2018, xu_understanding_2020}, requiring precise alignment of narrow beams between the base station (gNB) and user equipment (UE) \cite{tassi_modeling_2017}.

High delays associated with mmWave HOs have significant impacts throughout 
the entire protocol stack degrading quality of service~\cite{hassan_vivisecting_2022, narayanan_first_2020}. 
To determine when a handover is needed, the standard protocol requires users to pause data communication, scan neighboring cells, and report the results to the serving cell.
This process requires users to engage in an exhaustive search for the strongest signal from neighboring cells, resulting in service interruptions and performance degradation~\cite{mezzavilla_end--end_2018}. 
While existing works~\cite{wang2020demystifying} primarily focus on reducing beam searching overhead, 
the fundamental problem of maintaining reliable connectivity remains unexplored. 
Also, the HO decision relies on these measurement from users, which may be outdated or lost due to the rapidly fluctuating channel conditions. This can trigger flawed decisions, leading to ping-pong handovers and connection timeouts, impacting overall network performance.

This paper presents the design and implementation of \shortname{}, 
a programmable smart surface strategically deployed on a vehicle’s interior to address these challenges. 
\shortname{} acts as a programmable L1 extension of base stations that overcomes vehicle body blockage and makes the HO process transparent to users.
It achieves this through a new PHY-layer capability called \textit{dual-beam control}.
\shortname{} steers two incoming beams simultaneously in reflective and/or transmissive directions, with arbitrary power division between them. 
Using this ability, it offloads the HO task from individual user to the surface itself. This allows all in-vehicle users to maintain data connectivity while the network performs HO process on their behalf.
As illustrated in \cref{f:intro_use_cases}, we summarize three key innovations:

\begin{figure}[t]
\centering{}\textbf{Without \shortname{}:}\qquad{}
\qquad{}\textbf{With \shortname{}:}\\
\centering
\begin{subfigure}[b]{.9\linewidth}
\includegraphics[width=1\linewidth]{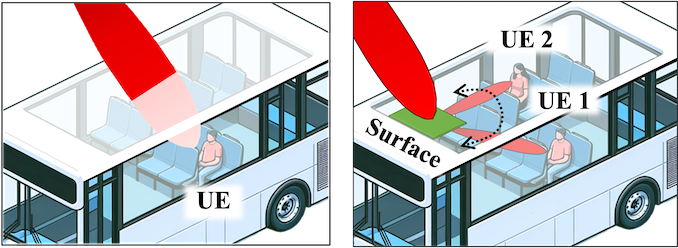}
\caption{Enhance in-vehicle mmWave penetration.}
\label{f:intro_use1}
\end{subfigure}
\begin{subfigure}[b]{.9\linewidth}
\includegraphics[width=1\linewidth]{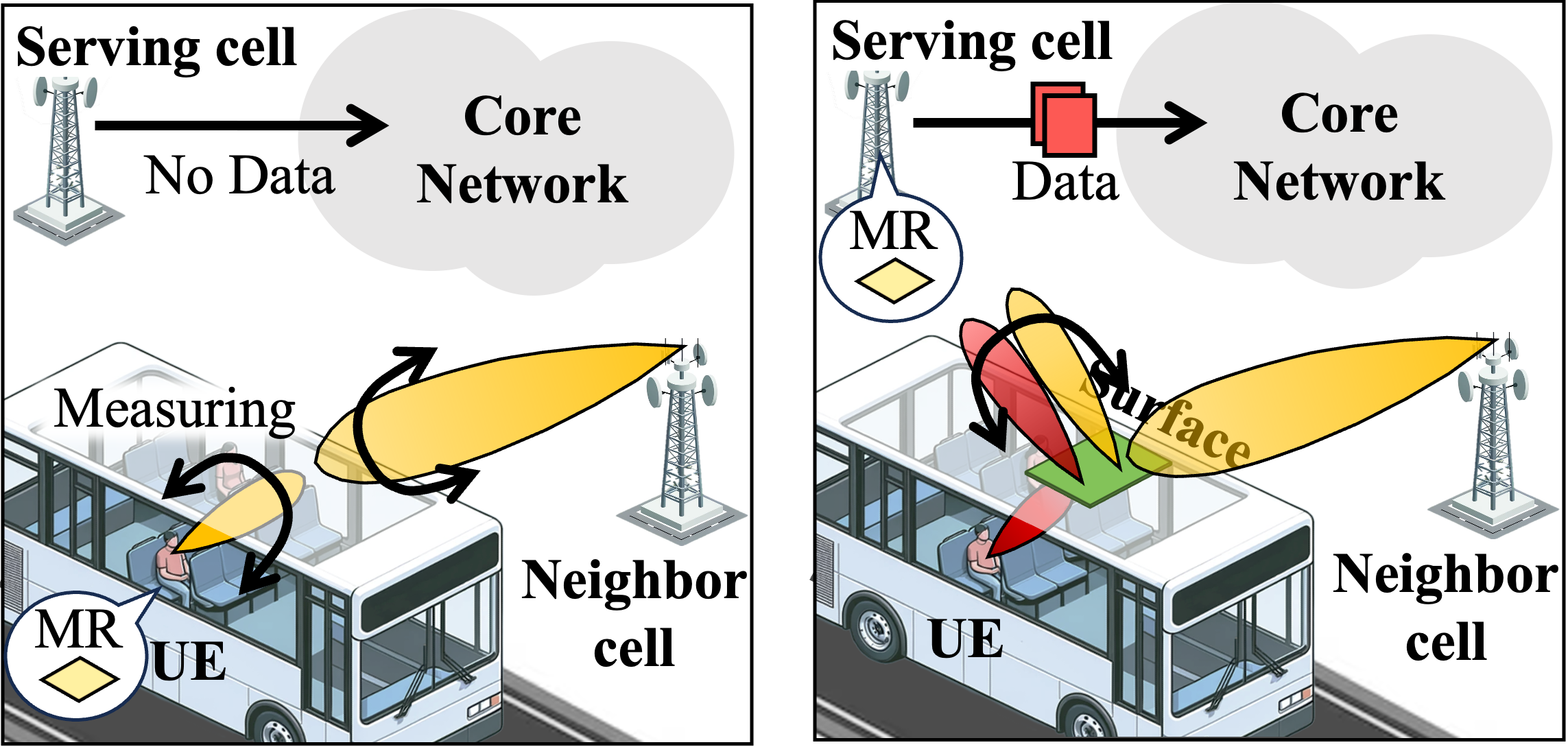}
\caption{Concurrent communications and cell scanning.}
\label{f:intro_use2}
\end{subfigure}
\begin{subfigure}[b]{.9\linewidth}
\includegraphics[width=1\linewidth]{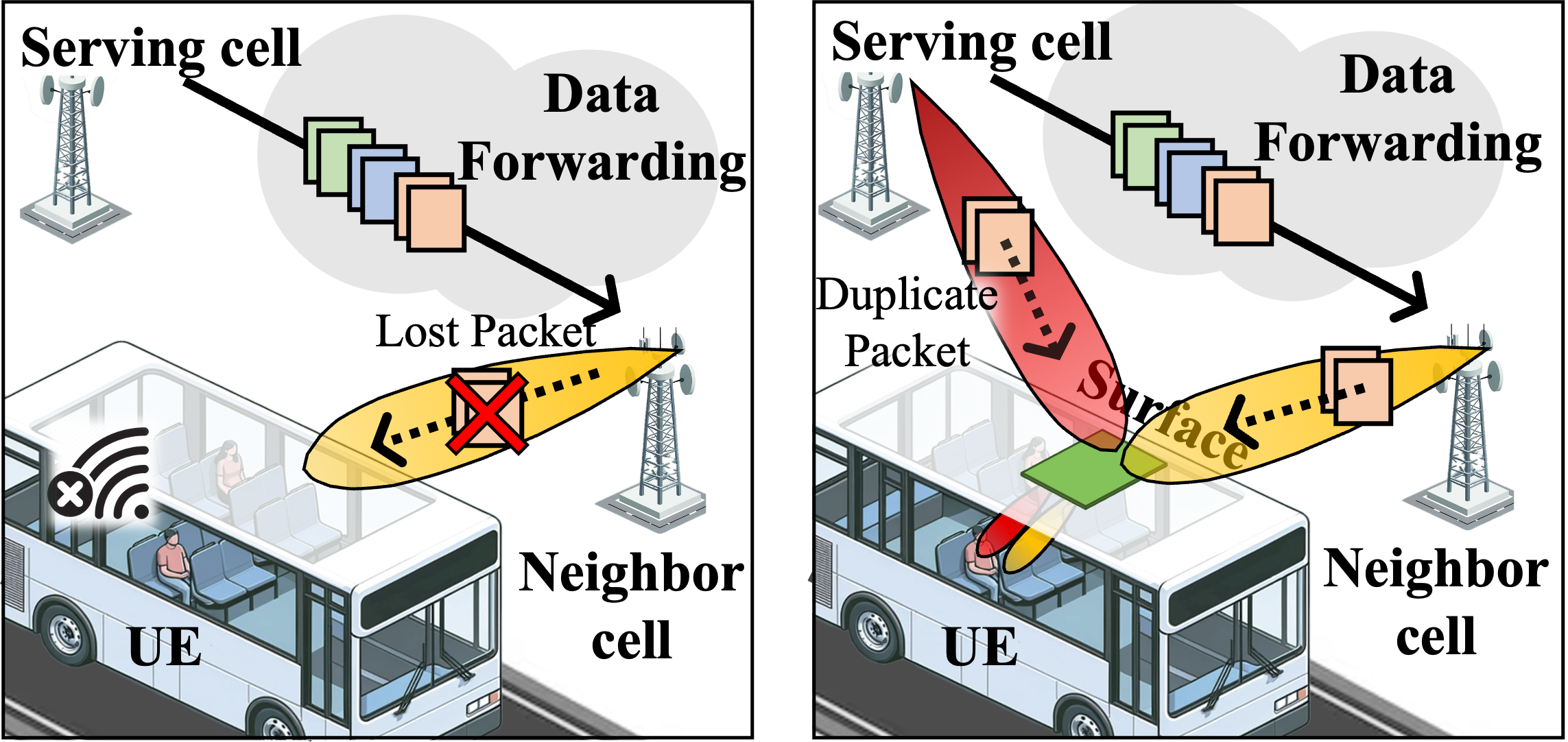}
\caption{Make-before-break handover.}
\label{f:intro_use3}
\end{subfigure}
\caption{\shortnames{} design innovations electronically
split, shape, and steer
mmWave transmissions in real time
to ensure seamless roadside mmWave networks.}
\label{f:intro_use_cases}
\end{figure}

\parahead{\textbf{1.}~Batched handover via shared in-vehicle coverage.} 
\shortname{} provides consistent, shared mmWave coverage to all users by steering outdoor signals into the vehicle (\cref{f:intro_use1}).
This also allows the network to batch the HO procedure for all in-vehicle users into a single operation, eliminating the overhead and contention of separate HOs.

\parahead{\textbf{2.}~Seamless handover with concurrent communication.} 
Innovating beyond existing surfaces that can only reflect or refract at one time \cite{cho_mmwall_2023}, 
\shortname{} uniquely steers two mmWave beams simultaneously in transmissive and reflective angles. 
It reflects neighboring cell’s broadcast signals to the serving cell while keeping the user data link through a transmissive path (\cref{f:intro_use2}).
This allows the serving cell to measure neighbor cells directly. 
By offloading the scanning burden from the user, our system eliminates service interruptions and significantly reduces UE power consumption associated to HOs, which consumes twice the energy compared to low-band HOs \cite{hassan_vivisecting_2022}.

\parahead{3.~Make-before-break handover.} 
At the moment of HO, \shortname{} maintains links from both the old and new cells. 
Users receive duplicate packets from both cells, reducing packet loss and retransmissions and enhancing reliability during cell transitions\footnote{
\textcolor{black}{To avoid collisions, nearby cells separate their respective signals in frequency and time. This allows the serving cell to listen to other cells during data communication and the user to receive signals from two cells.}}.

We have designed, implemented, and integrated the \shortname{} system into the COSMOS wireless testbed \cite{raychaudhuri_challenge_2020}. 
We have also implemented new PHY features on the testbed. 
We mount \shortname{} on the rear door of an SUV with multiple mobile users inside, 
communicating with three cells placed on the first floor of a lab facing the road. 
To evaluate TCP performance, we use a mmWave ns-3 simulator~\cite{polese2019end}. 
We modified the simulator to be trace-driven, fed with PHY traces collected from driving experiments.
In multi-user scenarios, our ns-3 results demonstrate that Wall-Street improves TCP throughput by up to 78\%, reduces RTT by up to 34\%, and cuts unnecessary HOs in half. 
Our microbenchmarks validate \shortnames{} in-vehicle coverage improvement, beam tracking, and dual-beam steering\footnote{\textcolor{black}{GitHub codes: \href{https://github.com/kunwooch/Wall-Street}{https://github.com/kunwooch/Wall-Street}}}\footnote{\textcolor{black}{Demonstration videos: \href{https://youtu.be/35iiBFYVkZY}{https://youtu.be/35iiBFYVkZY}}}.

\parahead{Contributions.} Our main contributions are as follows:
\begin{itemize}[leftmargin=*]
\item \textbf{A dual-link beam manipulation.} 
We design, build, and demonstrate the first mmWave smart surface capable of simultaneously and independently steering a transmissive link (into the vehicle) and a reflective link (between cells) with programmable power division. 
\item \textbf{Reliable make-before-break handover.} 
We realize a make-before-break handover via \shortname{}, which connects in-vehicle users with two cells. By receiving packets from both cells, our system improves reliability.
\item \textbf{Scalable, vehicle-centric handover.} 
We introduce an algorithm that makes a single, batched handover for all in-vehicle users, reducing signaling overhead. We leverage measurements from multiple users to estimate the cell-to-surface link quality and make a reliable handover decision.
\item \textbf{Real-world system implementation.}
We implemented the \shortname{} hardware and integrated it within the COSMOS testbed. We conducted driving experiments with a surface-equipped SUV and multiple gNB/UE nodes.
\end{itemize}

\section{Related Work}
\label{s:related}
\parahead{High-speed Mobility Management.}
Several works address sub-6 GHz vehicular mobility. Wi-Fi Goes to Town \cite{song_wi-fi_2017} provides continuous connectivity via rapid switching between roadside Wi-Fi APs. 
POLYCORN \cite{ni_polycorn_2023} optimizes traffic scheduling for high-speed railways but requires fixed trajectories,
making it unsuitable for unpredictable vehicular movement.
REM \cite{li_beyond_2020} uses multipath profiles instead of signal strength for HO decisions. 
However, these systems target sub-6 GHz bands, where signals penetrate obstacles easily with minimal cell search overhead. 
At mmWave, car bodies block signals completely and narrow beams lose alignment easily, demanding fundamentally different solutions.

\parahead{High-speed mmWave Connectivity.}
Measurement studies reveal severe mmWave mobility challenges.
\cite{hassan_vivisecting_2022} shows mmWave HOs occur every 0.13 km versus 0.4 km for low-band 5G, causing 2 Gbps throughput drops, 107\% higher latency, 5$\times$ more signaling overhead, and 2.4$\times$ higher UE battery drain. 
While \cite{wang2020demystifying} demonstrates efficient beam searching for vehicle-to-infrastructure (V2I) links, it addresses a fundamentally different problem. 
First, it treats the vehicle itself as a UE with roof-mounted antennas.
While external antennas bypass blockage, our measurements show that in-vehicle users suffer 70–80\% outage due to the vehicle body.
Second, the receiver in \cite{wang2020demystifying} does not perform directional beamforming, which naturally simplifies the search process.
Lastly, rather than optimizing search speed, our work prioritizes service continuity, maintaining UE data flow while the network seamlessly prepares and executes handovers.
\begin{table}
\footnotesize
\makeatletter
\newcommand*\ex[8]{#1&#2&#3&#4&#5&#6&#7&#8
}
\makeatother
\begin{threeparttable}
\caption{Comparison to existing works.}
\label{tab:features}
\begin{tabular}{@{}lc c c c c c c}
\toprule
  \multirow{1}{.8cm}{Related works} &  \multirow{1}{.3cm}{Freq. (GHz)} & \multirow{2}{.6cm}{Mobility support} & \multirow{1}{.4cm}{Steer-able} & \multirow{2}{.6cm}{\centering Tra/ Ref} &  \multirow{2}{.6cm}{\raggedleft Soft HO} & \multirow{2}{.5cm}{\centering Power (W)} & \multirow{2}{.5cm}{E2E impl.}\\
  \\
\midrule
\midrule

\ex{MiliMirror \cite{qian_millimirror_2022}}
{\cellcolor{green!25}60}{\cellcolor{red!25}Fixed}{\cellcolor{red!25}$\times$}{\cellcolor{red!25}R}
{\cellcolor{red!25}$\times$}{\cellcolor{green!25}\checkmark}{\cellcolor{yellow!25}-}\\

\ex{NR-Surface \cite{kim_nr-surface_2024}}
{\cellcolor{green!25}24}{\cellcolor{red!25}Human}{\cellcolor{green!25}\checkmark}{\cellcolor{red!25}R}
{\cellcolor{red!25}$\times$}{\cellcolor{green!25}\checkmark}{\cellcolor{green!25}\checkmark}\\

\ex{mmWall \cite{cho_mmwall_2023}}
{\cellcolor{green!25}24}{\cellcolor{red!25}Human}{\cellcolor{green!25}\checkmark}{\cellcolor{yellow!25}T or R}
{\cellcolor{red!25}$\times$}{\cellcolor{green!25}\checkmark}{\cellcolor{yellow!25}-}\\

\ex{V2I RIS~\cite{chen2021robust,alsenwi2022intelligent,tian2022optimizing}}
{\cellcolor{green!25}28}{\cellcolor{green!25}Vehicle}{\cellcolor{green!25}\checkmark}{\cellcolor{red!25}R}
{\cellcolor{red!25}$\times$}{\cellcolor{green!25}\checkmark}{\cellcolor{red!25}$\times$}\\


\midrule

\ex{Repeater \cite{movandi, pivotal}}
{\cellcolor{green!25}28}{\cellcolor{green!25}Vehicle}{\cellcolor{green!25}\checkmark} {\cellcolor{red!25}T}
{\cellcolor{red!25}$\times$}{\cellcolor{red!25}>20}{\cellcolor{green!25}$\checkmark$}\\

\midrule
\ex{\textbf{Wall-Street}}
{\cellcolor{green!25}26}{\cellcolor{green!25}Vehicle}{\cellcolor{green!25}\checkmark}{\cellcolor{green!25}T \& R}
{\cellcolor{green!25}\checkmark}{\cellcolor{green!25}0.48}{\cellcolor{green!25}\checkmark}\\
\bottomrule
\end{tabular}
\label{t:comparison}
\end{threeparttable}
\end{table}

\parahead{Mobile mmWave Repeaters.}
Commercial mmWave repeaters for homes~\cite{pivotal} and vehicles~\cite{movandi} use external antennas to receive signals, amplify, and retransmit them to users inside. 
This architecture relies on digital processing and active RF chains, resulting in a power consumption of 20–40 W. This far exceeds \shortname{}'s 160 \micro W power\footnote{With two COTS DACs, the total power consumption is 0.48 W. We refer to Sec.~\ref{s:discussion} for a detailed power analysis.}.
Also, controlling two mmWave beams simultaneously would require repeaters to add more RF chains, further increasing power.
Repeaters also incur higher search complexity (\cref{s:design:attach}).
Finally, rooftop-to-WiFi relays are suboptimal as they introduce latency and limit mmWave’s multi-Gbps throughput to WiFi speeds.

\parahead{Reconfigurable Intelligent Surface (RIS).}
RISs controls mmWave signals using programmable metamaterials that directly phase-shift incoming waves.
As seen in \cref{tab:features},
\cite{qian_millimirror_2022} uses passive RISs to reflect signals in fixed directions but unable to support mobile users. 
More advanced reflective RISs use switches~\cite{tan_enabling_2018}, PIN diodes~\cite{yang_two-dimensional_2022, gros_reconfigurable_2021, gros_design_2023}, or varactor diodes ~\cite{wolff_continuous_2023, kim_nr-surface_2024, cho_mmwall_2023} for beam steering. 
While \cite{kim_nr-surface_2024} integrates 5G protocols, they are limited to reflection mode, making them unsuitable for in-vehicle coverage.
\cite{cho_mmwall_2023} introduces the ability to switch between transmission and reflection but requires time-multiplexing. \shortname{} supports simultaneous transmission and reflection with flexible power division.

While roadside RIS deployments~\cite{chen2021robust,alsenwi2022intelligent,tian2022optimizing} help mitigate environmental blockage but require extensive infrastructure and cannot solve vehicle body penetration. 
\shortname{}, mounted directly on vehicles, moves with users to ensure continuous in-vehicle coverage and reliable handovers.

\section{Primer: Standalone Handover}
\label{s:primer}

Standalone (SA) base stations exchange handover signaling via the \emph{Xn interface}.
The process has three phases: \textit{Preparation}, \textit{Execution}, and \textit{Completion}.

\parahead{(1)~Preparation phase.}
The network relies on UE measurements to trigger handovers. 
The serving cell periodically requests UEs to pause ongoing transmissions to measure the Reference Signal Received Power (RSRP) of neighboring gNBs. 
The UE performs an exhaustive search and sends a Measurement Report (MR) containing neighboring gNB IDs and RSRP values back to the serving gNB. 
A handover is typically triggered by \emph{event A3} 
\cite{deng_mobility_2018, xu_understanding_2020}, 
which occurs when the neighboring gNB’s RSRP ($M_n$) exceeds the serving cell’s RSRP ($M_s$) by a hysteresis threshold $H$ ($M_n  >  M_s + H$).
To prevent ping-pong HO from signal fluctuations, HO executes when A3 holds true for the time-to-trigger (TTT) duration.







\parahead{(2)~Execution phase.}
The \textsf{serving gNB} sends a handover request to the selected
\textsf{target gNB} via the Xn interface. 
Upon acknowledgment, the serving gNB commands the UE to detach from itself and synchronize with the target using Random Access (RACH), while forwarding buffered data to the target gNB. 
Because this transition often occurs at the service boundary, 
the newly established link suffers from packet loss and retransmissions.

\parahead{(3)~Completion phase.}
The target gNB requests path switching to the Core and signals the serving gNB to release the UE context via the Xn interface.
Poor HO decisions often lead to ping-pong handovers or TCP connection termination.


\section{Design}
\label{s:design}
This section presents the design of \systemname{}, including the surface design (\cref{s:design:hardware}) and handover design (\cref{s:design:ho}).

\subsection{Wall-Street Hardware}
\label{s:design:hardware}
\shortname{} has three key contributions: 
(1) \shortname{} operates at 26 GHz with dual-link steering and power division; 
(2) \shortname{} provides an adaptive control layer that is transparent to the user. Unlike conventional relays requiring active UE coordination, \shortname{} externalizes UE control, offloading mobility management to the infrastructure;
(3) by manipulating links in the analog domain, \shortname{} minimizes the energy and latency of digital processing required by active relays (\textit{e.g.,} receiving, decoding, encoding, and re-transmitting).

\begin{figure}
\includegraphics[width=1\linewidth]{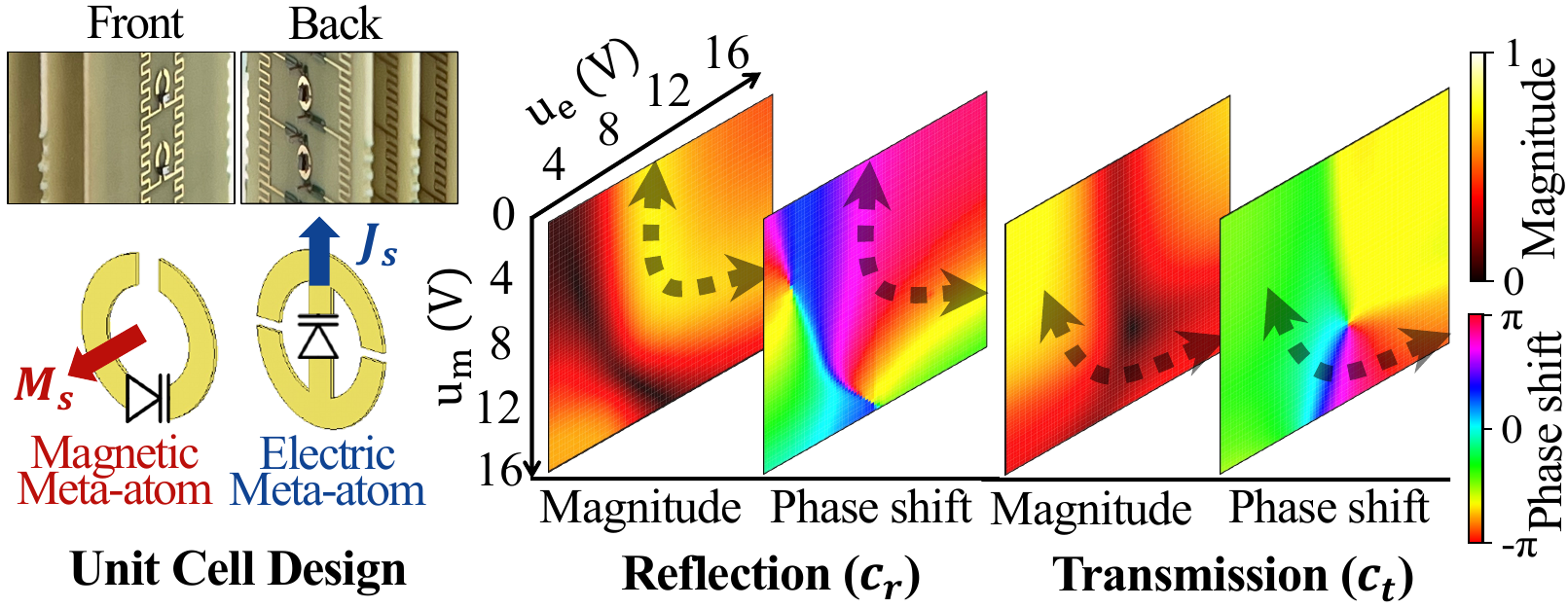}
\caption{(\textit{left}) \shortname{} consists of over 4,000 paired electric and magnetic meta-atoms separated by a dielectric substrate. Biasing voltage on varactor diodes modulates current oscillation, creating EM field discontinuities; (\textit{right}) VNA measured reflection and transmission response at 26 GHz, varying with biasing voltages ($u_m$, $u_e$).}
\label{f:huygens}
\end{figure}

\parahead{Surface Design.}
\shortname{} is a 26 GHz Huygens metasurface (HMS) that supports both reflection and transmission with the gain of $29$ dB. 
HMS consists of orthogonal electric and magnetic meta-atoms facing each other across a dielectric substrate (shown in \cref{f:huygens} on left), creating discontinuities in EM fields~\cite{cho_mmwall_2023}.
This structure can manipulate the magnitude and phase of the incident field in both reflective and transmissive directions.
Each meta-atom contains a low-power voltage-controlled capacitor.
By adjusting the bias voltages applied to the magnetic and electric meta-atoms \((u_m,u_e)\), we can change the complex reflection and transmission coefficients of the incoming signal.
\Cref{f:huygens} (right) shows Vector Network Analyzer (VNA) measurements of \shortname{}'s response to different voltage combinations.
The marked regions (black arrows) indicate where we achieve arbitrary phase shifts from \(0\) to \(2\pi\) while maintaining high efficiency in reflection or transmission.

\parahead{Dual-Link Beam Steering.}
\shortname{} simultaneously steers one beam in transmission and another in reflection with adjustable power splitting.
The challenge is to find an optimal configuration that jointly satisfies these multiple objectives, achieving two steering angles with a desired power ratio.

A straightforward approach is to spatially partition the surface into two - for example, assigning 80\% of the area to the transmissive link and 20\% to the reflective link. A distinct phase gradient is then applied to each partition to steer the two beams independently.
While it is computationally fast, hard partitioning raises undesirable sidelobes and reduces gain for both links.
Instead, we use soft partitioning, which optimizes the entire surface to serve both objectives simultaneously. This approach finds a single voltage configuration that maximizes gain in both directions while maintaining the desired power split and suppressing sidelobes.

We pre-compute these configurations using a genetic algorithm and store them in a lookup table (codebook) for real-time access.
To populate this codebook, we find the optimal voltages 
$\left\{u_{m,n}, u_{e,n}\right\}$ 
that maximize the combined array factor for a target pair of angles ($\theta_1$, $\theta_{2}$) and power split $\alpha$:
\begin{align}
\max_{\{u_{m,n},u_{e,n}\}}~ \Bigg| \sum_{n=0}^{N-1} \Big(
\underset{\text{Transmissive link at } f_1}{ \underbrace{\sqrt{1-\alpha}\, c_{t,n}\, e^{-j\phi_{t,n}}}} + 
\underset{\text{Reflective link at } f_2}{ \underbrace{\sqrt{\alpha}\, c_{r,n}\, e^{-j\phi_{r,n}}}}
\Big) \Bigg|^2
\label{eq:array_factor_hms}
\end{align}
where $\phi_k$ is the ideal phase shift for steering to angle $\theta_k$.
The terms \(c_{t,n}\) and \(c_{r,n}\) denote the complex transmission and reflection coefficients of the $n$-th meta-atom, derived from S-parameter measurements in \cref{f:huygens} (\textit{right}).
The resulting codebook \(\Theta\), indexed by \((\theta_t,\theta_r,\alpha)\), is used during real-time operation described in \cref{s:design:ho}. 
For dual-transmission, the reflective term is replaced by a second transmissive term. 
For single-beam steering, the second term is removed entirely.

\begin{figure}
\begin{subfigure}[b]{.49\linewidth}
    \includegraphics[width=1\linewidth]{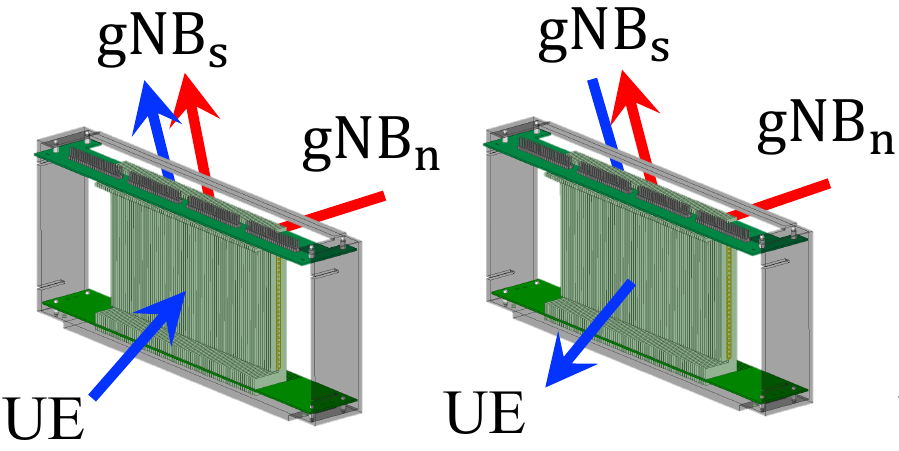}
    \caption{Cell Search (UL/DL)}
    \label{f:design:split1}
\end{subfigure}
\begin{subfigure}[b]{.49\linewidth}
    \includegraphics[width=1\linewidth]{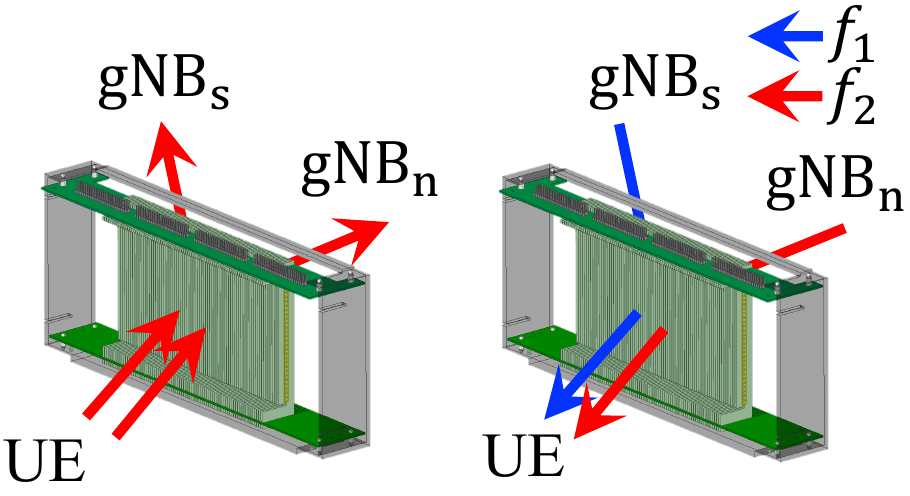}
    \caption{Make-Before-Break (UL/DL)}
    \label{f:design:split2}
\end{subfigure}
\caption{\textbf{Wall-Street’s multi-beam operational modes:} (a) dual-transflective steering for cell search and (b) dual-transmissive steering for make-before-break. Blue and red beams operate at different frequencies.}
\label{f:eval:beamsplitting}
\end{figure}


\begin{figure*}
    \begin{subfigure}[b]{.13\linewidth}
    \centering
    \includegraphics[width=1\linewidth]{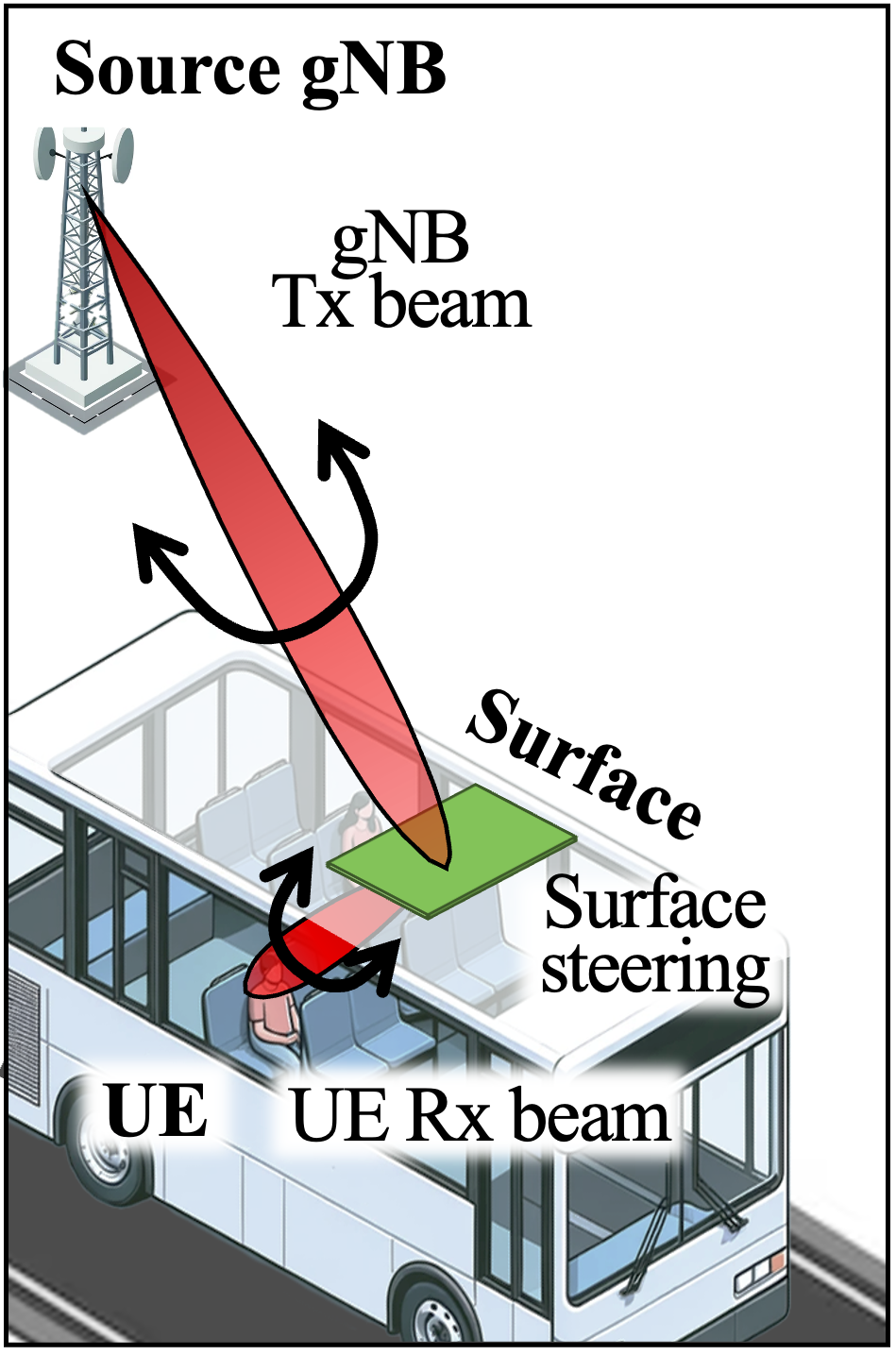}
    \caption{Initial attachment.}
    \label{f:overall:initial}
    \label{f:overall:attach}
    \end{subfigure}
    \begin{subfigure}[b]{.345\linewidth}
    \centering
    \includegraphics[width=1\textwidth]{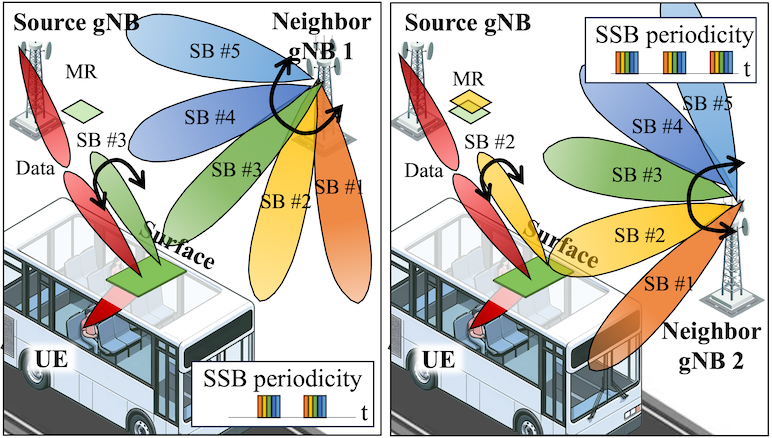}
    \caption{Handover Preparation: Measurement of candidate 
    gNB~1 and gNB~2.}
    \label{f:overall:prep}
    \end{subfigure}
    \begin{subfigure}[b]{.34\linewidth}
    \centering
    \includegraphics[width=1\textwidth]{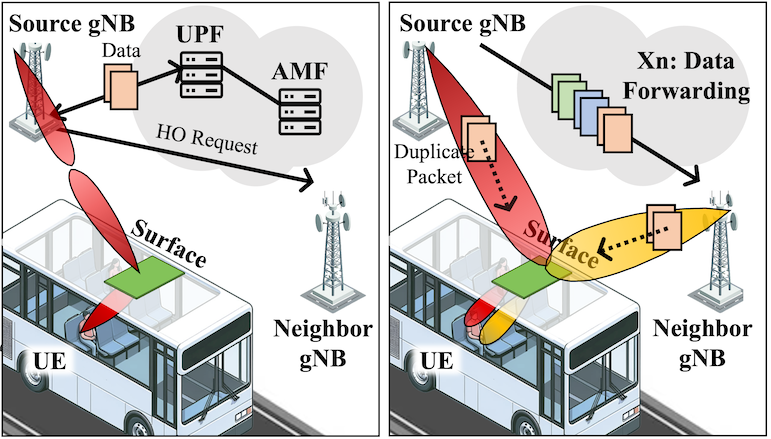}
    \caption{Handover Execution: Make-before-break handover.}
    \label{f:overall:ho}
    \end{subfigure}
    \begin{subfigure}[b]{.17\linewidth}
    \centering
    \includegraphics[width=\textwidth]{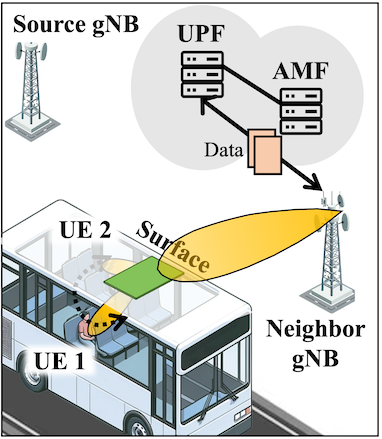}
    \caption{Handover Completion.}
    \label{f:overall:compl}
    \end{subfigure}
    \caption{\textbf{System overview:}~\textbf{(a)}~Initial 
    UE-RAN attachment (\cref{s:design:attach}); 
    \textbf{(b)}~preparation using transflective surface power measurement (\cref{s:design:ho:prep}); 
    \textbf{(c)}~beam combining to enable make-before-break dual-cell connectivity 
    (\cref{s:design:ho:execute});
    \textbf{(d)}~handover completion (\cref{s:design:ho:complete}). Each gNBs use different frequencies, and the source gNB controls the surface.}
    \label{f:overall_design}
\end{figure*}

\shortname{}'s dual-link beam steering, which independently controls two angles across a wide $\pm70\degree$ range, provides two handover functions, as illustrated in \cref{f:eval:beamsplitting}:
\textbf{(1) neighbor cell search }(\cref{f:design:split1}): \shortname{} maintains a transmissive link between the serving gNB and UE (blue) while reflecting neighbor cell's broadcast signals back toward the serving gNB (red) for cell search, and
\textbf{(2) make-before-break handover} (\cref{f:design:split2}): \shortname{} steers both the serving and target gNB's signals to the UE.
Furthermore, we leverage surface's angular reciprocity, where the beam path is identical for uplink and downlink. This allows for instantaneous TDD switching without surface reconfiguration. 
In contrast, active repeaters must switch between receive and transmit beam patterns, introducing additional latency

\begin{figure}
\begin{subfigure}[b]{.635\linewidth}
\centering
\includegraphics[width=1\linewidth]{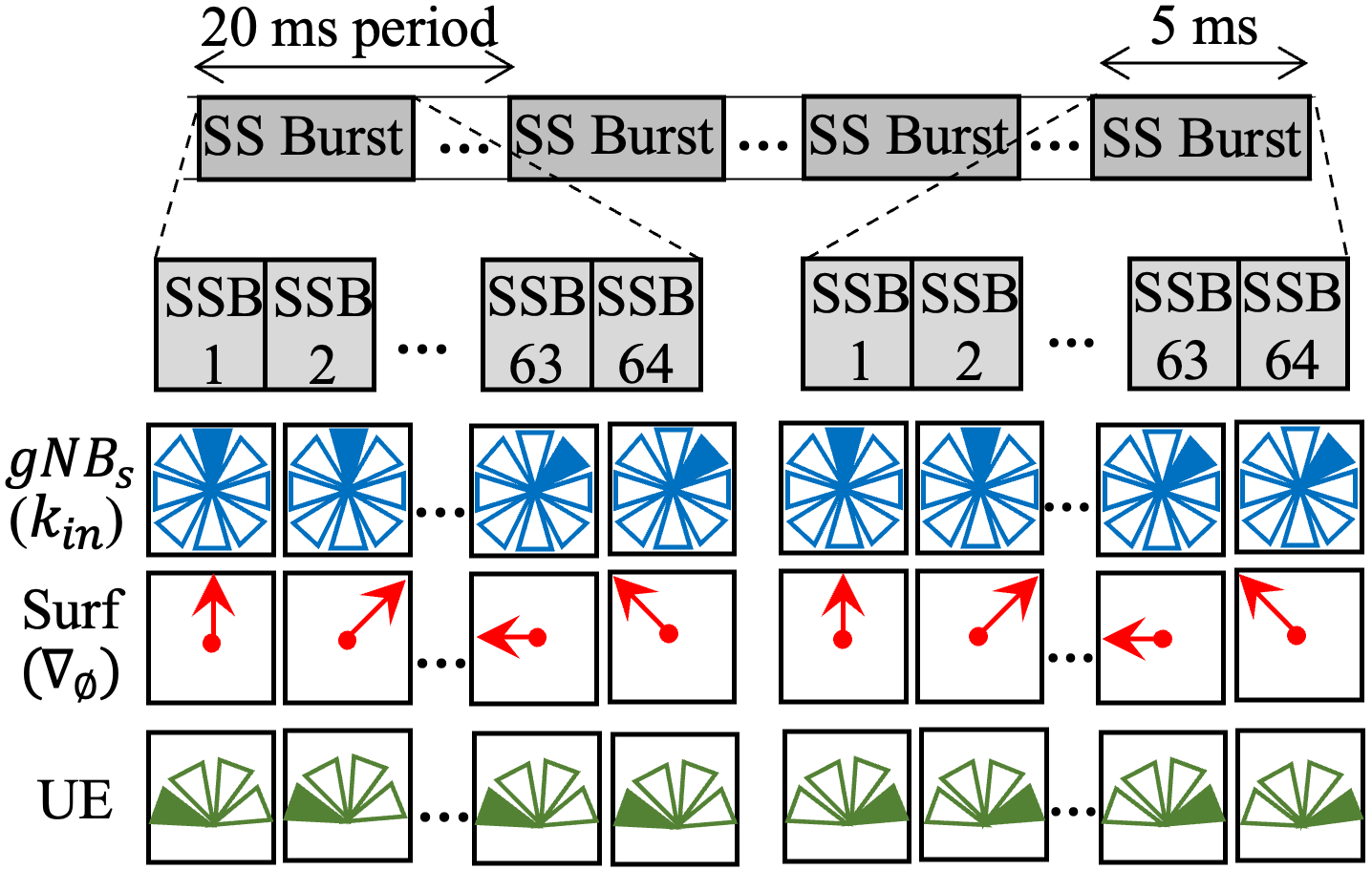}
\caption{Initial attachment.}
\label{f:wp1_beam}
\end{subfigure}
\centering
\begin{subfigure}[b]{.355\linewidth}
\includegraphics[width=1\linewidth]{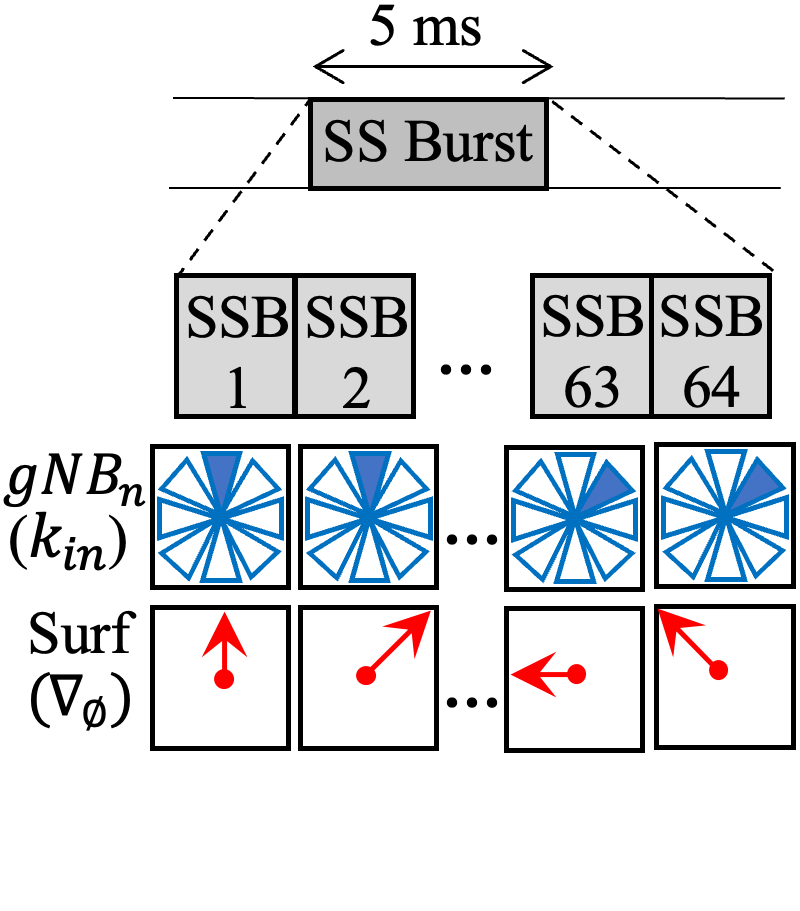}
\caption{Cell scanning.}
\label{f:wp2_beam}
\end{subfigure}
\caption{\textbf{Alignment process:}~initial attachment and cell scanning during HO preparation.}
\label{f:wallstreet_beam_acquisition}
\end{figure}

\subsection{Wall-Street Handover Design}
\label{s:design:ho}
This section describes \shortname{}'s HO process, which is divided into: initial attachment, preparation, decision, execution, and completion, as detailed in 
\cref{f:overall_design}.
First, to ensure uninterrupted connectivity, we offload the neighbor cell measurement task from users to the surface (\cref{s:design:ho:prep}). 
Second, to improve efficiency, we batch the HO for all in-vehicle users into a single operation  (\cref{s:design:ho:decision}). 
Finally, to maximize reliability, we establish a make-before-break connection, linking users to the new cell before disconnecting from the old cell (\cref{s:design:ho:execute}).

\parahead{System Assumptions.}
Our design operates under three architectural assumptions:
\begin{itemize}
    \item Nearby gNBs transmit their broadcast signals on different frequency resources. This is a standard deployment practice to prevent signal collisions. We leverage this practice for both handover preparation and execution. 
    \item The serving gNB can simultaneously measure reflected signals from a neighbor while maintaining communication with the UE on a separate frequency.
    \item The serving gNB controls \shortname{} via a sub-6 GHz control channel. Instead of accessing the surface’s low-level codebooks, gNBs signals high-level beam steering angles. The surface’s onboard microcontroller has the hardware-specific codebook and locally maps angles into the voltage configurations.
\end{itemize}
We further discuss protocol integration, 3GPP compatibility, and methods to reduce system complexity in \cref{s:discussion}.

\subsubsection{Initial Attachment}
\label{s:design:attach}
\label{s:design:normalop}
As illustrated in \cref{f:wp1_beam}, the initial attachment requires a coordinated beam sweep to align the gNB, \shortname{} (operating in a single-beam transmission mode), and the UE. 
Every $20$ ms period, the gNB broadcasts a $5$ ms burst containing 64 synchronization signal blocks (SSBs), which are transmitted across 8 different gNB Tx beam directions.
For each SSB, \shortname{} refract them into $8$ angles, testing different paths into the vehicle. 
For each burst, the UE holds its Rx beam and measures the power.
The UE tests a different beam direction for each burst. 
After four bursts (one per UE Rx direction) in $80$ ms, the UE reports the best alignment and connects with the gNB.

\subsubsection{Handover Preparation}
\label{s:design:ho:prep}
The serving gNB periodically triggers \shortname{} to enter a dual-beam mode for neighbor cell measurement. 
The goal is to measure neighbor cell RSRP without interrupting data sessions, as shown in \cref{f:overall:prep}. 
Since the refractive link between the serving gNB, \shortname{}, and the UE is already established, the transmissive angle $\theta_t$ is known.
\shortname{} then uses its pre-computed codebook indexed by $(\theta_t, \theta_r, \alpha)$ to cycle through 8 reflective angles $\theta_r$, 
and neighbor gNBs transmit SSBs in their standard beam patterns.
Since the serving gNB already knows its direction toward \shortname{},
it measures these reflected SSBs on a separate channel to prevent interference with the active data link.
The power split $\alpha$ allocates less power to maintain the data connection while assigning the remaining power to the reflective link. 
As illustrated in \cref{f:wp2_beam}, this search takes one 5 ms burst.
This allows the serving gNB to measure the neighbor RSRPs on behalf of the UEs without their involvement or service interruption.

\subsubsection{Handover Decision.}
\label{s:design:ho:decision}
\begin{figure}
\centering
\includegraphics[width=0.55\linewidth]{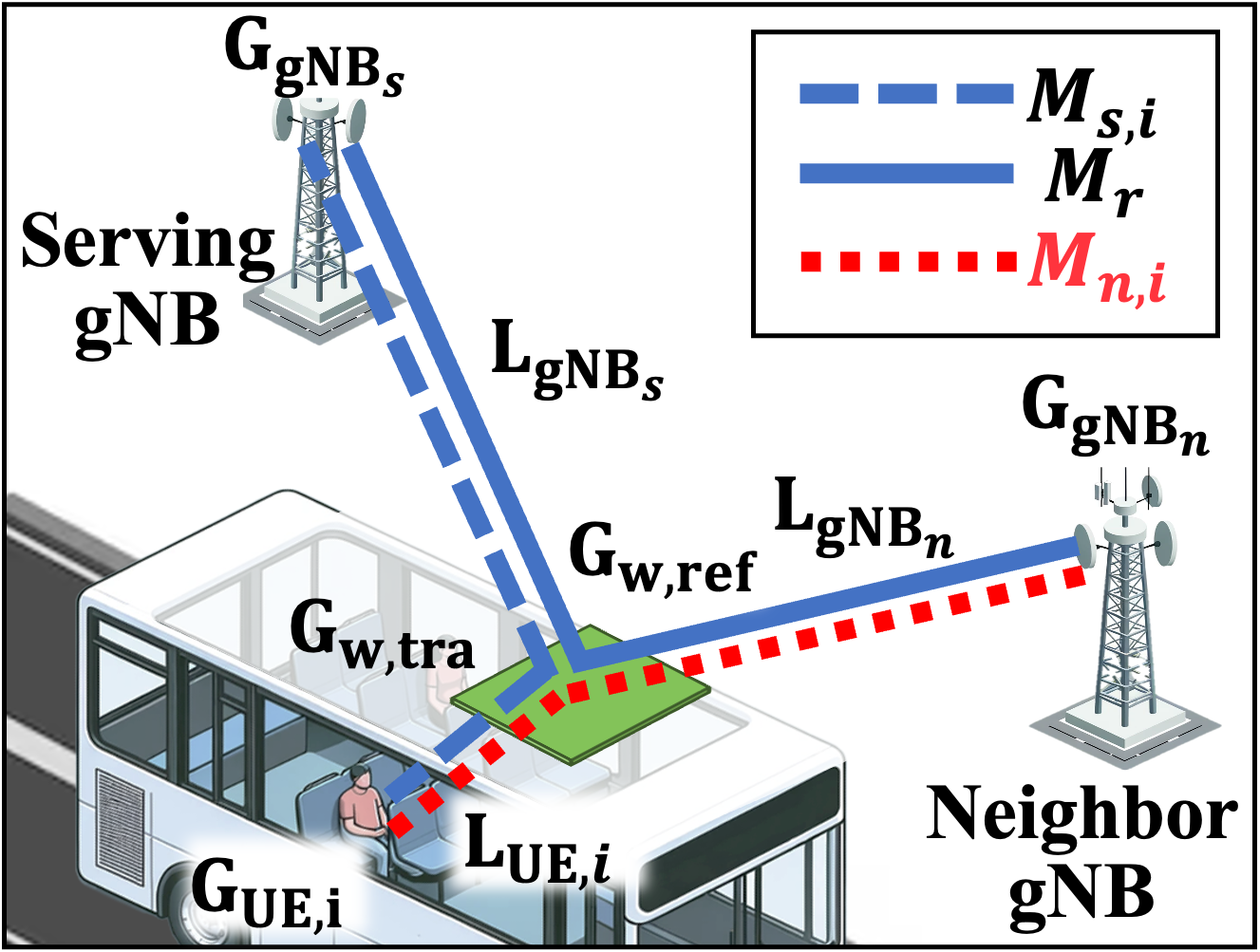}
\caption{\textbf{Measurement power translation:}~Inferring 
measurement report power readings based on measurements reflected from \shortname{}.}
\label{f:rsrp_cal}
\end{figure}

After measuring neighbor cells, the serving gNB determines if new handover will improve signal quality for all UEs inside the vehicle. 
Standard A3 HO triggers when the neighbor RSRP ($M_{n,i}$) exceeds the serving RSRP ($M_{s,i}$) by a threshold $H$ ($M_{n,i} > M_{s,i} + H$) for $i$-th UE. 
However, this per-user HO could lead to conflicting decisions for different UEs.
Also, the serving gNB measures the reflected signal $M_r$ instead of $M_{n,i}$. 
As illustrated in \cref{f:rsrp_cal}, we simplify the serving, neighbor, and reflected RSRP as:
\begin{align}
\label{eq:m_s}
M_{s,i} &= (G_{\mathrm{gNB}_s} + L_{\mathrm{gNB}_s}) + G_{\mathrm{w,tra}} + (G_{ue,i} + L_{ue,i}) \\
M_{n,i} &= (G_{\mathrm{gNB}_n} + L_{\mathrm{gNB}_n}) + G_{\mathrm{w,tra}} + (G_{ue,i} + L_{ue,i}) \\
M_r     &= (G_{\mathrm{gNB}_n} + L_{\mathrm{gNB}_n}) + G_{\mathrm{w,ref}} + (G_{\mathrm{gNB}_s} + L_{\mathrm{gNB}_s})
\label{eq:m_r}
\end{align}
where $G_{gNB,s}$ and $G_{gNB,n}$ denote the serving gNB and neighbor gNB gain, respectively. 
Similarly, $L_{gNB,s}$ and $L_{gNB,n}$ represent the path loss between the surface and the serving and neighbor gNBs.
The serving gNB directly measures $M_{s,i}$ from each UE and $M_r$ from \shortname{}. 
It also knows surface gains $G_{w,tra}$ and $G_{w,ref}$ from the codebook and the UE gain $G_{ue,i}$.\footnote{In 3GPP, the gNB sends Downlink Control Information (DCI) to send power control commands to the UE.}
The UE-specific path loss $L_{ue,i}$ remains unknown.

Instead of using gNB-UE links, our approach is to make one HO decision for the entire vehicle based on the estimated gNB-surface links.
We redefine the serving link as $X_{s} = G_{gNB,s} + L_{gNB,s}$ and the neighbor link as $X_{n} = G_{gNB,n} + L_{gNB,n}$.
Then, the HO decision for all $K$ users becomes $X_n > X_s + H$.
Our goal is to estimate $X_s$ and $X_n$ from available measurements.
To do so, we employ a two-slot measurement with bounding algorithm described next.

\parahead{Two-slot measurements.}
We use \shortname{}'s dynamic power allocation to estimate $X_s$ and $X_n$ accurately. 
It allocates power between reflection ($\alpha_j$) and refraction ($1-\alpha_j$) in each time slot $j$. 
In Slot 1, we increase reflection to measure accurate $M_r$ (already measured in \cref{s:design:ho:prep}), while in Slot 2, we increase refraction to accurately measure $M_{s,i}$.
Based on \cref{eq:m_s} and \cref{eq:m_r}, we define the per-slot models as:
\begin{align}
    &M_r^{(j)} = X_n + X_s + G_{w,ref}(\alpha_j) \\
    &M_{s,i}^{(j)} = X_{s} + G_{w,tra}(1-\alpha_j) + G_{ue,i} + L_{ue,i}^{(j)} 
\end{align}

\parahead{Bounding algorithm. }
We use measurements $X_{s,i}$ from multiple UEs during Slot 2 to tightly bound $X_s$. 
We first subtract the known surface and UE gains from our measurements:
\begin{align}
    &S_1 = M_{r}^{(1)} - G_{w,ref}(\alpha_1) = X_s + X_n    \label{eq:slot1} \\
    &S_{2,i} = M_{s,i}^{(2)} - G_{w,tra}(1-\alpha_2) - G_{ue,i} = X_s + L_{ue,i}^{(2)}
    \label{eq:slot2}
\end{align}
The path loss within the vehicle is physically bounded as $L_{min} \leq L_{ue,i} \leq L_{max}$. 
We can bound $X_s$ based on each UE measurement as $X_s \in [S_{2,i} - L_{max}, S_{2,i} - L_{min}]$. 
By combining measurements from all UEs, we tighten this bound as:
\begin{align*}
    X_s \in [\underset{i \in K}{\text{max}} (S_{2,i} - L_{max}), \underset{i \in K}{\text{min}} (S_{2,i} - L_{min})]=[LB_s, UB_s]
\end{align*}
Based on \cref{eq:slot1}, we can write our decision metric as $\Delta = X_n - X_s = S_1 - 2X_s$, so the bounds for our metric becomes
$\Delta_{min} = S_1 - 2UBs$.
We trigger HO if $\Delta_{min} \geq H$. Otherwise,
HO does not occur to avoid unnecessary HOs.
This conservative approach prevents rapid ping-pong HOs caused by transient signal fluctuations.

\begin{figure*}
\begin{subfigure}[b]{0.6\linewidth}
\includegraphics[width=1\linewidth]{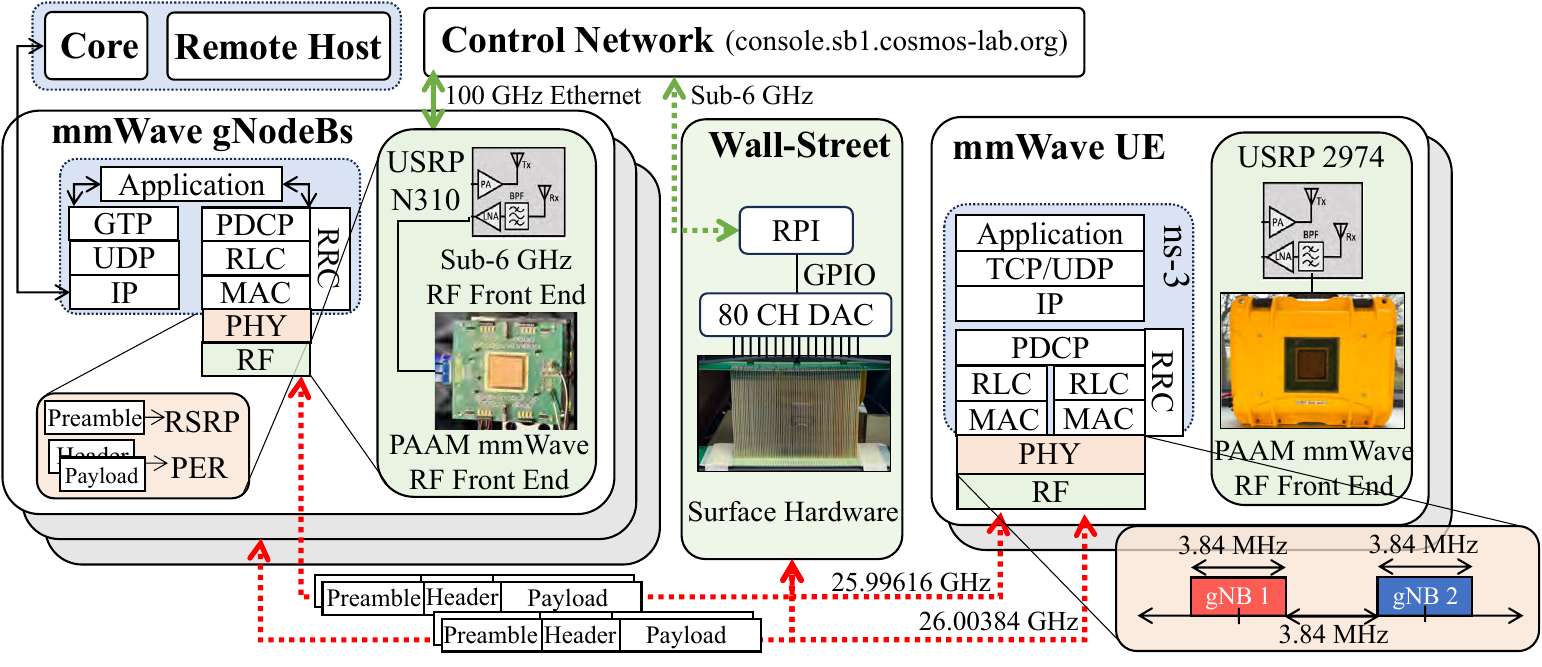}
\end{subfigure}
\begin{subfigure}[b]{0.39\linewidth}
\centering
\includegraphics[width=1\linewidth]{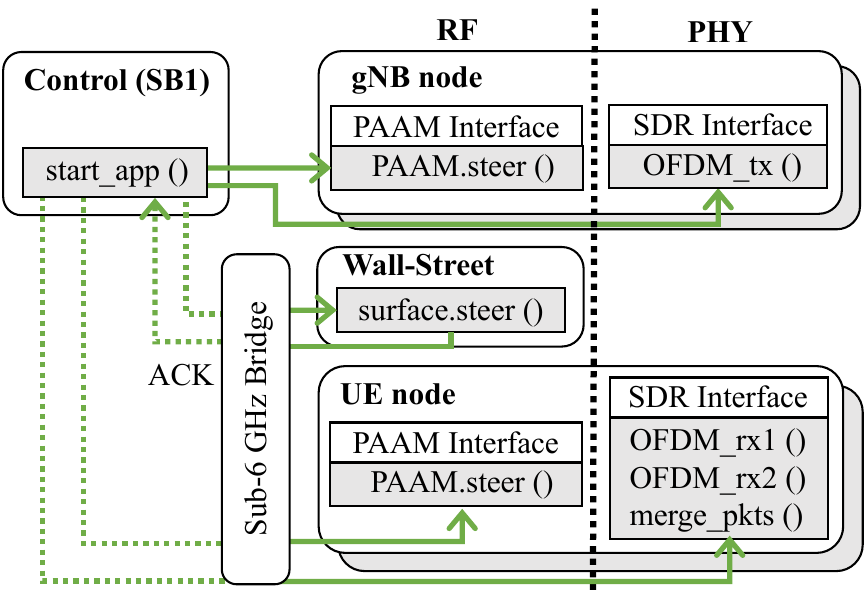}
\end{subfigure}
\caption{\textbf{\systemname{}-COSMOS testbed integration:} we integrate
\systemname{} with COSMOS USRP-based gNBs and UE (\textit{left}).
 We illustrate a simplified control flow of \shortname{}-integrated COSMOS testbed (\textit{right}).}
\label{f:cosmos_int}
\end{figure*}

\subsubsection{Handover Execution}
\label{s:design:ho:execute}
The serving gNB initiates the HO by sending a \textsf{HO~Req} message to the target gNB. 
Upon receiving the \textsf{HO~ACK}, the serving gNB triggers \shortname{} to enter a make-before-break mode and sends \textsf{HO command} to UEs. 
UEs synchronize with the target gNB via RACH and maintain two active protocol stacks using the 3GPP-standardized Dual Active Protocol Stack (DAPS)~\cite{3gpp_5g_2021}.  
It establishes new Signaling Radio Bearers (SRBs) for control-plane signaling, including the \shortname{} control, with the target cell while suspending the serving cell SRBs.
For the user plane, a common Packet Data Convergence Protocol (PDCP) entity is created to serve both links, allowing UEs to exchange data with both gNBs on different channels.

At the physical layer, \shortname{} combines downlink beams from both gNBs and splits the uplink beam towards both gNBs.
No additional beam searching is required for this step as the optimal angles for the serving link ($\theta_1$) and the target link ($\theta_2$) are already known from HO preparation.
To ensure reliable packet delivery, both the serving and target gNBs transmit duplicate data packets to UEs, 
where the PDCP layer handles reordering and de-duplication. 
During make-before-break, \shortname{} can directly measure both \(M_s\) and \(M_n\) to check A3 condition in real time.
A3 conditions are continuously evaluated. If the condition is not met, the system reverts to the serving cell.



\subsubsection{Handover Completion}
\label{s:design:ho:complete}

The target cell completes the HO by
sending a \textsf{UE release} to the serving cell and \textsf{HO completion} to \shortname{}. 
Upon completion, 
\shortname{} maximizes transmission power to the new gNB.
If the HO fails, \shortname{} instead maximizes power to the old gNB so that UEs do not restart from the initial attachment phase.
UEs also reverts to the old gNB configuration and 
reactivates old cell SRBs for control-plane signaling.
We note that a single HO execution simultaneously transfers all UEs within the same vehicle, reducing control-plane signaling.

\section{Implementation}
\label{s:impl}
\Cref{f:cosmos_int} illustrates our implementation of \shortname{}.
It consists of
\textbf{(1)}~a surface hardware and control unit; 
\textbf{(2)}~integration with the COSMOS
testbed, implementing new PHY features; and 
\textbf{(3)}~an
mmWave ns-3 simulation~\cite{polese2019end} for evaluating TCP performance, 
which was modified to be trace-driven and 
fed with PHY traces collected from COSMOS testbed. 


\subsection{\shortname{} Hardware}
\begin{figure*}
    \begin{subfigure}[b]{.295\linewidth}
    \centering
    \includegraphics[width=1\linewidth]{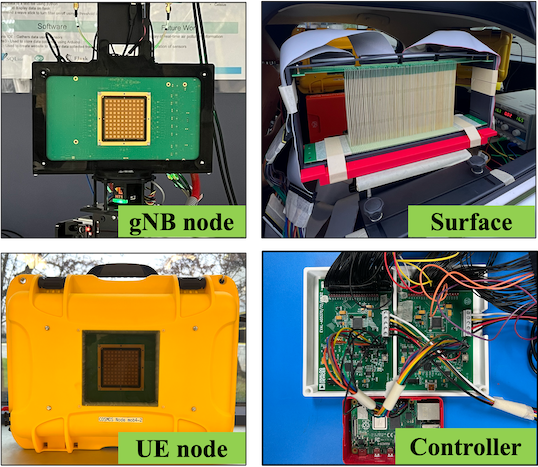}
    \caption{\shortname{} hardware}
    \label{f:impl:hardwares}
    \end{subfigure}
    \begin{subfigure}[b]{.172\linewidth}
    \centering
    \includegraphics[width=1\linewidth]{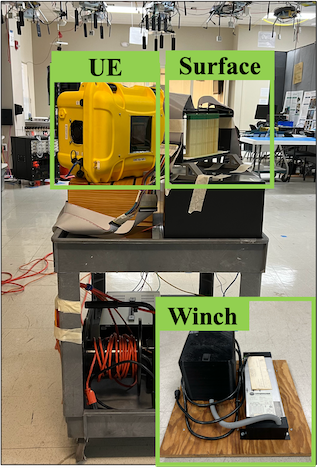}
    \caption{Indoor setup}
    \label{f:impl:indoors}
    \end{subfigure}
    \begin{subfigure}[b]{.52\linewidth}
    \centering
    \includegraphics[width=1\linewidth]{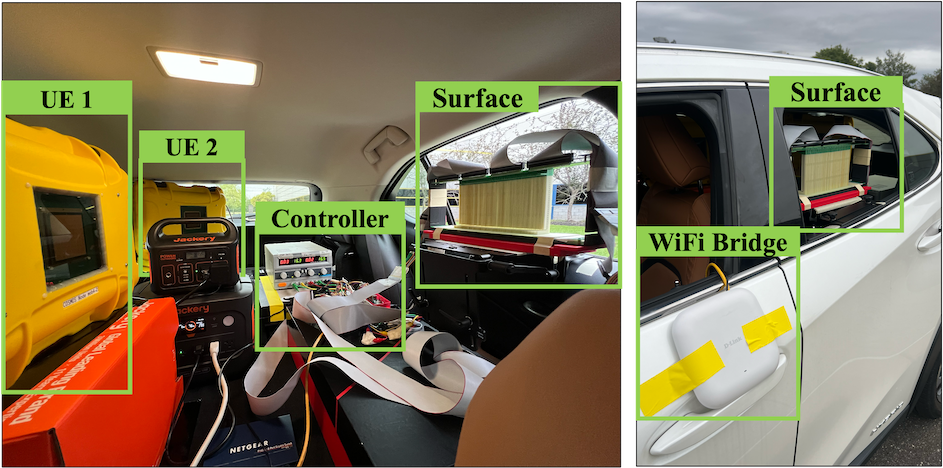}
    \caption{Outdoor in-vehicle multi-UE setup}
    \label{f:impl:outdoors}
    \end{subfigure}
    \caption{\textbf{\shortnames{} hardware implementation and testbed:}~showing \textbf{(a)}~individual
    hardware components, \textbf{(b)}~indoor experimental UE-side setup,
    and \textbf{(c)}~outdoor in-vehicle UE-side configuration.}
    \label{f:impl}
\end{figure*}

The \shortname{} surface is composed of 76 boards, each consisting of 28 co-located magnetic and electric unit cells as shown in \cref{f:impl:hardwares}.
Two 40-channel AD5370 16-bit DACs provide independent control of the unit cells on each board,
with each channel supplying a variable 0 to 16 V control voltage. 
To expedite the steering process, one channel supplies the voltage to two adjacent boards, enabling two DACs to independently control all boards.
The two DACs are each connected to the Serial Peripheral 
Interface (SPI) of a Raspberry Pi (RPI) through GPIO.
The RPI listens for control signals from the COSMOS testbed control network, known as sandbox1 (SB1), via sub-6 GHz. 
Upon receiving a signal, it sends a command to the DACs, which then apply the appropriate voltage levels. 
These voltage levels are determined from a pre-stored angle-to-voltage codebook. 
For multi-beams, we construct codebook for different power ratios between two beams ($\alpha=0.25, 0.5,0.75$).
The speed of the \shortname{} hardware has 
been optimized to 0.2~ms for real-time experiments.

\begin{figure}
\centering
 \begin{subfigure}[b]{.44\linewidth}
    \centering
    \includegraphics[width=1\linewidth]{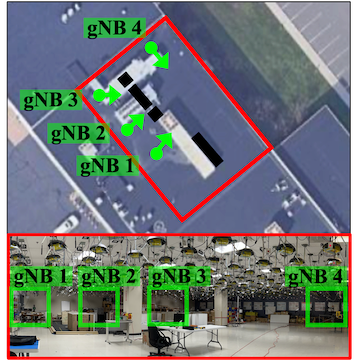}
    \caption{Indoor setup}
    \label{f:impl:testbed1}
    \end{subfigure}
    \begin{subfigure}[b]{.51\linewidth}
    \centering
    \includegraphics[width=1\linewidth]{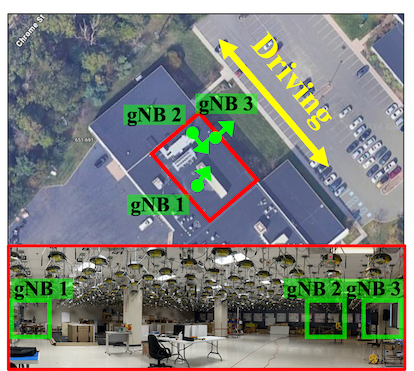}
    \caption{Outdoor setup}
    \label{f:impl:testbed2}
    \end{subfigure}
\caption{\textbf{Testbed setup and map:}~we 
deploy the gNBs on the first floor 
of a lab facing an outdoor parking lot. The vehicle is 
approximately $20$ to $30$~m~from the gNBs. The arrow indicates the gNB direction facing at 0\degree{}, and each node sweeps the beam from -60\degree{} to 60\degree{}.}
\label{f:impl:outdoor_setup}
\end{figure}

\subsection{COSMOS Testbed}
The COSMOS testbed comprises software-defined radios (USRPs) integrated with the \href{https://wiki.cosmos-lab.org/wiki/Hardware/SubSystems/IBM#IBM28-GHzPhasedArrayBoardForCOSMOS}{IBM mmWave Phased-Array Antenna Module (PAAM)} frontends (\cref{f:cosmos_int}). The PAAM features 4 tiled RFICs, each with 32 TRx phase-shifting elements, supporting a total of 64 antennas. 
The PAAM offers full access to beamforming control and latency as low as 10s of nanoseconds for low-latency MAC and hybrid beamforming.
As shown in \cref{f:impl:hardwares}, our experimental setup has two types of PAAM devices: 
two stationary PAAM nodes integrated with USRP N310s mounted on remotely controllable XY tables, and 
three portable PAAM nodes integrated with USRP 2974s. 
For indoor scenarios, we utilize four gNBs: two stationary PAAMs (gNB 1 and 4) with a 35 dB gain and two portable PAAMs (gNB 2 and 3) with a 31 dB gain. 
In multi-UE outdoor scenarios, we use three gNBs: two stationary PAAMs (gNB 1 and 2) with a 65 dB gain and one portable PAAM (gNB 3) with a 31 dB gain, 
with the remaining portable PAAMs serving as UEs\footnote{The node gains are defined by the COSMOS testbed. We refer the reader to \href{https://www.cosmos-lab.org/technology/mmwave/}{https://www.cosmos-lab.org/technology/mmwave/} for the details.}. 
The location of UEs remain the same with and without \shortname{}.
Since all nodes must be connected to the same control network (SB1), 
we use a Wi-Fi bridge operating at sub-6 GHz to send the control commands to outdoor UEs and the surface.

\subsubsection{PHY implementation on COSMOS}
We developed new features on the PHY layer of the COSMOS testbed, including OFDM packet transmission and reception on two different channels, extraction of RSRP and Packet Error Rate (PER) from the packet, and IBM PAAM and Wall-Street beam control. 
Periodic bursts of OFDM packets are sent from two PAAMs in different frequency bands with 3.84 MHz separation, 
and the receiver simultaneously decodes packets from both transmitters by operating at a larger bandwidth. 
The receiver extracts RSRP from the reference signal in the preamble, the packet sequence number from the header, and the PER from the cyclic redundancy check (CRC). 
During make-before-break, the serving and target gNBs transmit identical packets, and 
the UE decodes packets on both channels simultaneously and compute the combined PER. 
A C++ script in SB1 sends control commands to the gNB and \shortname{} to follow the handover procedure (\cref{f:cosmos_int} on right).

\subsubsection{NS-3 simulation}
The COSMOS testbed currently supports mmWave PHY-layer experiments but lacks the full upper-layer stack for TCP traffic. We modified an existing mmWave NS-3 simulator to be trace-driven and fed with the PHY-layer COSMOS traces for TCP analysis. 
Specifically, we modified its standalone Hard Handover (HH) scheme (we refer to \cite{polese2019end} for details). 
We made major modifications to its PHY and RRC modules to support trace-based operations and handover logic. The PHY module runs a continuous data transmission cycle. 
Throughout this cycle, the spectrum PHY module continuously updates the data SINR, syncing with timestamps and SINRs from COSMOS data transmission traces. 
Running independently from the data transmission cycle, a scanning cycle is triggered at a defined measurement report periodicity. 
During this cycle, the PHY module loops through beam combinations in a 5-ms burst, retrieving SINR values from COSMOS beam scanning traces. Once scanning is complete, the results are reported to the RRC module to evaluate A3 event conditions. The serving cell’s RRC module initiates handover for all UEs when 
$\Delta_{min} \geq H$.

\section{Evaluation}
\label{s:eval}
We present microbenchmarks on beam patterns, coverage improvement, and dual-link capabilities of \shortname{} (\cref{s:eval:microbenchmarks}).
We then conduct field studies to collect traces from the COSMOS testbed under driving scenarios. Finally, we input these PHY traces into the ns-3 simulator to analyze TCP throughput and latency performance (\cref{s:eval:e2e}). 



\subsection{Methodology}
\label{s:eval:methodology}

We evaluate \shortname{} in both controlled indoor (single-UE) and real-world outdoor (multi-UE) driving settings.
For indoor settings, there are four gNB nodes and one UE node in a $20 \times 30$~m lab, as shown in \cref{f:impl:testbed1}.
To test the UE and \shortname{} under mobility, we place them on a cart moving in a constant speed of approximately 1~km/h.
For outdoor settings, there are three gNB nodes on the first floor of a lab facing a parking lot. 
We mount the surface on the rear window of a SUV vehicle and deploy two UE nodes.
UE 1 is on the rear seat (opposite side from surface), and UE 2 is on the cargo area, as shown in \cref{f:impl}. 
The vehicle drives at speeds ranging from 5 to 15 km/h. 
Both the surface and UEs are battery-powered, with UE positions remaining the same with and without \shortname{}.
Since the gNB nodes use stationary PAAMs that cannot be relocated, our outdoor driving trajectory is limited to 30 m, as depicted in \cref{f:impl:outdoor_setup}.

\begin{figure}
\centering
\includegraphics[width=.9\linewidth]{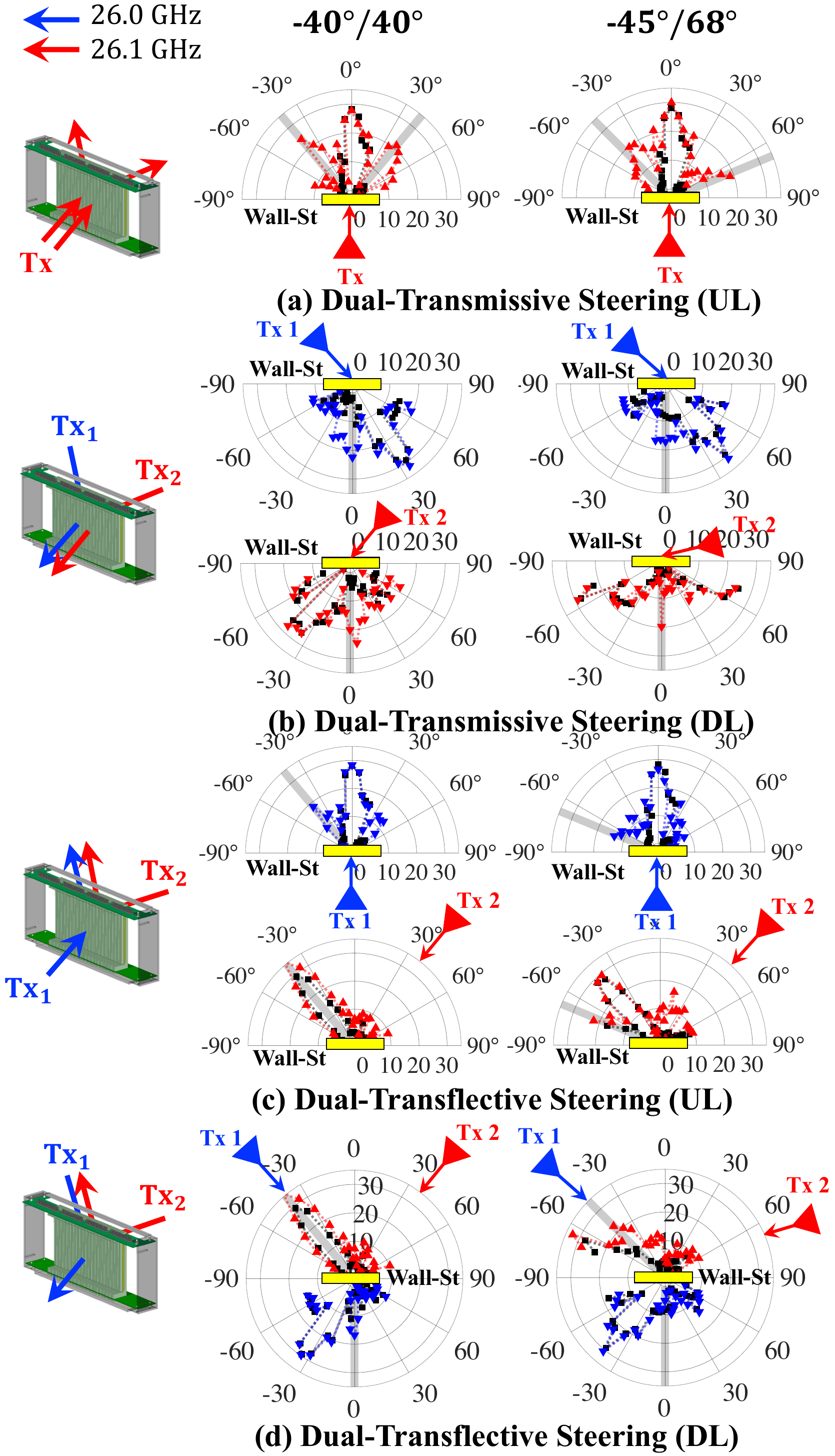}
\caption{\textbf{Beam patterns for four steering modes}.
Red/blue $\triangle$ show measurements at 26.0/26.1 GHz. 
$\blacksquare$ show baseline (surface turned off).
Gray lines indicate groundtruth angles.
\textit{Left}: -40\textdegree{}/40\textdegree{} steering; 
\textit{right}:~-45\textdegree{}/68\textdegree{} steering. 
}
\label{f:beam_microbench}
\end{figure}

We implement the standard SA protocol as our baseline.
For both \shortname{} and baseline, the hysteresis parameter $H$ of A3 event is set to $10$ dB with TTT of $150$ ms and the measurement report periodicity of $160$ ms. For the baseline, a total duration of every measurement report takes less than $20$ ms with three neighbor gNBs. 
Also, we set the maximum in-vehicle distance to $1$ m for $L_{min}$.
Our results combine (1) PHY measurements, including RSRP and packet error rate, from the COSMOS testbed with real vehicle trajectory, and
(2) TCP evaluation using ns-3 simulator driven by PHY traces. 
As RSRP values from the COSMOS testbed serve as relative signal strength indicators defined by the testbed and do not directly map to absolute dBm scales, we applied a 10 dB calibration offset to all traces, aligning them with the absolute dBm scale required by the ns-3 simulator.

\subsection{Microbenchmarks}
\label{s:eval:microbenchmarks}
We evaluate \shortnames{} far-field dual-beam patterns measured with a spectrum analyzer, in-vehicle signal coverage, and dual-beam performance at the physical layer. 

\parahead{Far-field beam patterns.}
\Cref{f:beam_microbench} demonstrates Wall-Street's beam steering capabilities in four configurations: 
dual-transmissive mode where both beams pass through the surface (\cref{f:beam_microbench} (a-b)), 
and transflective mode where one beam transmits through while another reflects (\cref{f:beam_microbench} (c-d)). 
Each configuration shows both uplink and downlink operations, validating its bidirectional functionality.

We evaluate two beam angle pairs (-40\textdegree/40\textdegree, -45\textdegree/68\textdegree).
To measure beam patterns, we place Tx and Rx horn antennas two meters from \shortname{} and sweep the Rx from -70\textdegree to 70\textdegree.
Two transmitters operate on separate frequencies, 26.0 GHz (blue $\triangle$) and 26.1 GHz (red $\triangle$), for simultaneous beam measurement. 
Gray lines indicate the groundtruth steering angles.
In the absence of an anechoic chamber, a significant portion of signal power diffracts around the surface. We quantify this background leakage by setting Wall-Street to a uniform 0V state—effectively acting as a specular mirror ($\blacksquare$).
We normalize all measurements to the spectrum analyzer’s noise floor (0 dB reference).


\begin{figure}
\centering
\includegraphics[width=1\linewidth]{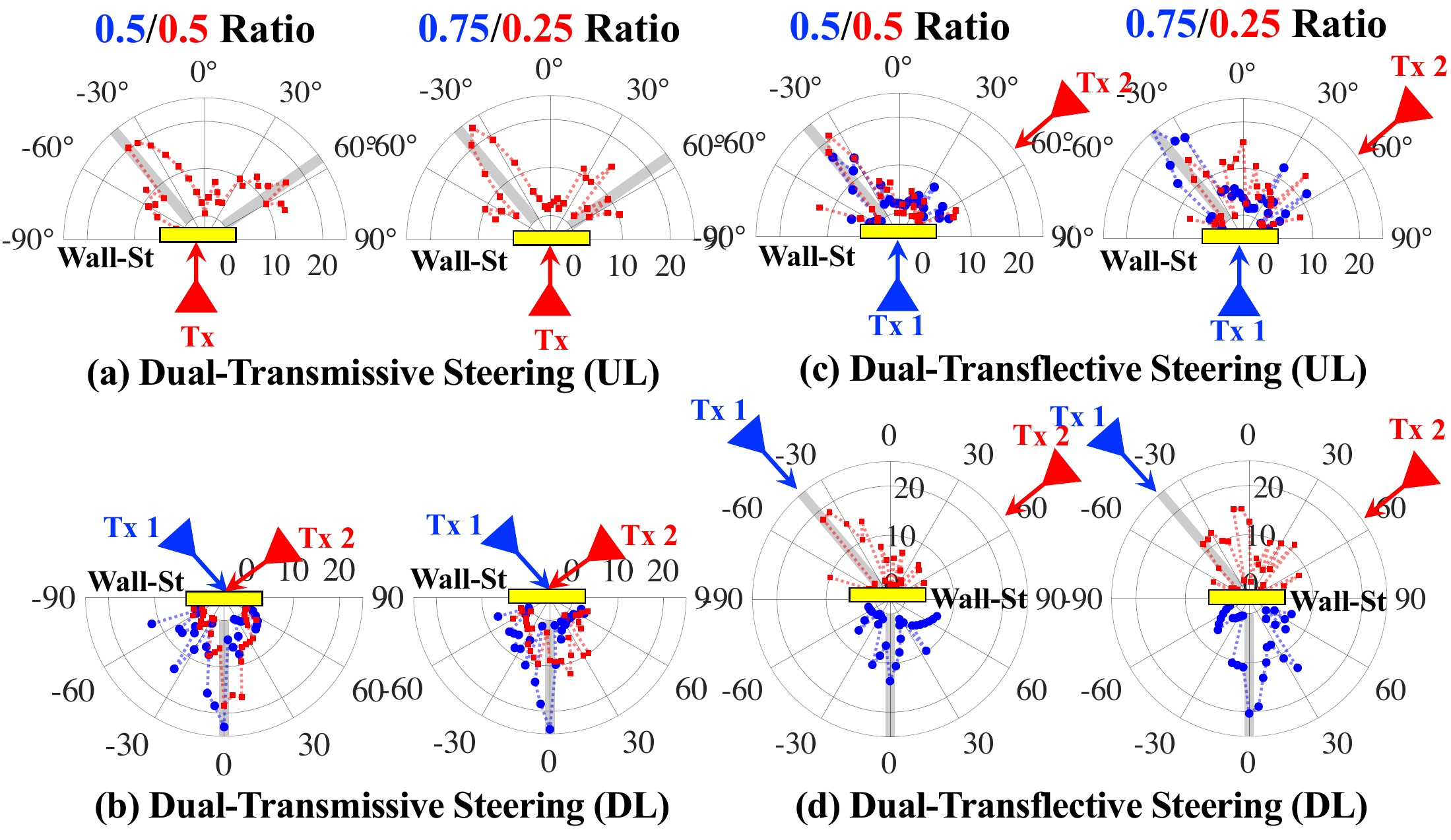}
\caption{\textbf{Power allocation} between beams at -40\textdegree/55\textdegree{}, normalized to baseline. Balanced (50/50) vs. asymmetric (75/25) ratio. Gray lines indicate groundtruth angles.}
\label{f:eval:power_ratio}
\end{figure}

We observe a 20–25 dB gain at ground-truth steering angles. This gain persists except when the steered beam coincides with the natural specular reflection angle, as seen at -40\textdegree/40\textdegree{} in \cref{f:beam_microbench}(c) and \cref{f:beam_microbench}(d), where the baseline ($\blacksquare$) itself is high.
This gain naturally decreases at wider steering angles. 
We use identical voltage configurations for both uplink and downlink (\textit{e.g.,} \cref{f:beam_microbench}(a) and \cref{f:beam_microbench}(b)), confirming angular reciprocity that allows fast UL/DL switching during handovers.


\parahead{Beam power ratio.}
\Cref{f:eval:power_ratio} shows dynamic power allocation between two beams at -40\textdegree/55\textdegree{}. 
All measurements are normalized against the baseline (surface off) for visibility.
Shifting from balanced (50/50) to asymmetric (75/25) ratio increases the first beam (blue) while decreasing the second beam (red) at groundtruth angles.
Both beams maintain groundtruth angles and only relative power changes.

\begin{figure}
\includegraphics[width=.9\linewidth]{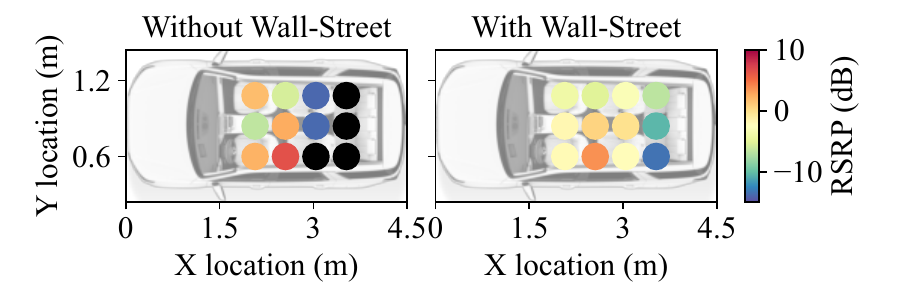}
\caption{In-vehicle mmWave coverage. $\bullet$:~no signal.}
\label{f:eval:coverage}
\end{figure}

\parahead{In-vehicle mmWave coverage.}
To evaluate Wall-Street’s ability to mitigate vehicle body blockage, we mounted the surface on the rear window with a gNB located 30 m away. For each interior location, we determined the maximum achievable RSRP through an exhaustive beam search.
As shown in \Cref{f:eval:coverage}, nearly 50\% of the vehicle interior experiences complete signal outage without \shortname{}.
In contrast, Wall-Street eliminates these outages entirely and provides a signal strength improvement of over 12 dB.


\begin{figure}
\begin{subfigure}[b]{.92\linewidth}
\includegraphics[width=1\linewidth]{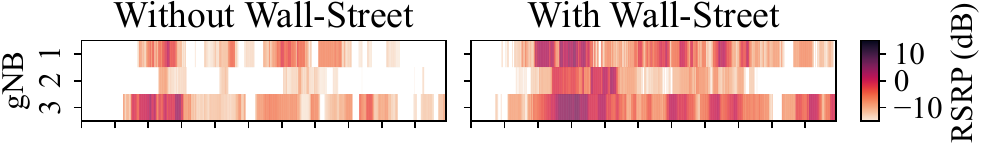}
\includegraphics[width=1\linewidth]{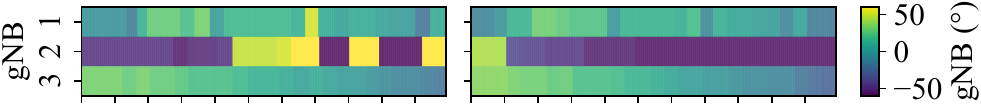}
\includegraphics[width=1\linewidth]{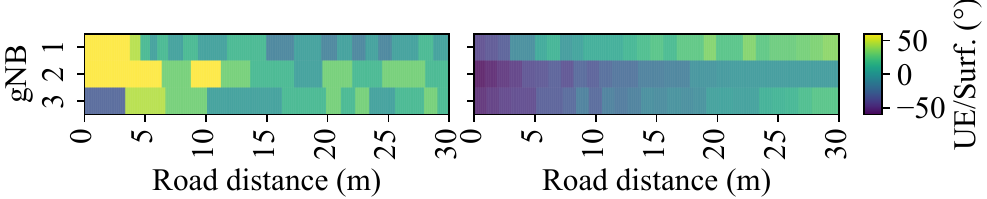}
\caption{Outdoor coverage for UE 1}
\label{f:eval:outdoor_rsrp_ue1}
\end{subfigure}
\begin{subfigure}[b]{.92\linewidth}
\includegraphics[width=1\linewidth]{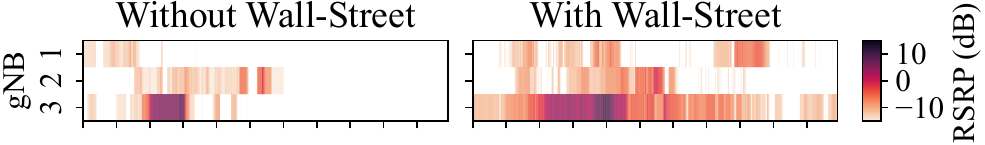}
\includegraphics[width=1\linewidth]{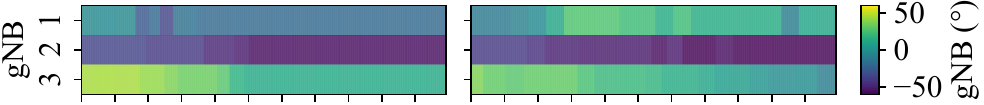}
\includegraphics[width=1\linewidth]{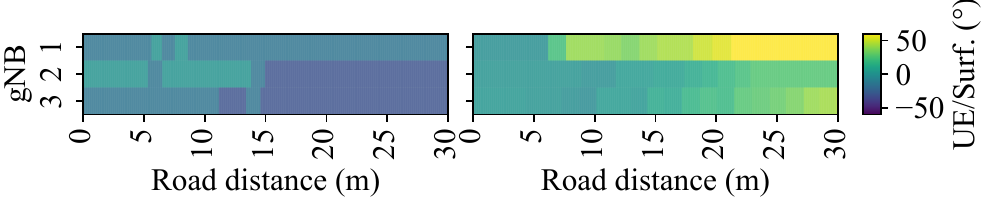}
\caption{Outdoor coverage for UE 2}
\label{f:eval:outdoor_rsrp_ue2}
\end{subfigure}
\caption{\textbf{Signal strength and beam angles during outdoor driving experiments} (\textit{upper:} RSRP; \textit{middle:} corresponding gNB beam angles; \textit{lower:} UE or surface angles).}
\label{f:eval:outdoor_rsrp}
\end{figure}

\parahead{Beam tracking under mobility.}
\Cref{f:eval:outdoor_rsrp} shows RSRPs and corresponding beam angles for gNB and UE/surface as the vehicle moves along a 30-meter outdoor path for UE 1 (\cref{f:eval:outdoor_rsrp_ue1}) and UE 2 (\cref{f:eval:outdoor_rsrp_ue2}). 
Overall, \shortname{} provides significantly higher signal strength.
This becomes particularly evident for UE 2. Without \shortname{}, UE 2 loses signals 70-80\% of the time, despite the presence of three gNBs. In contrast, \shortname{} ensures continuous connectivity from either gNB 1 or gNB 3 throughout the entire path.


\begin{figure}[t]
\centering
\begin{subfigure}[b]{.47\linewidth}
\includegraphics[width=1\linewidth]{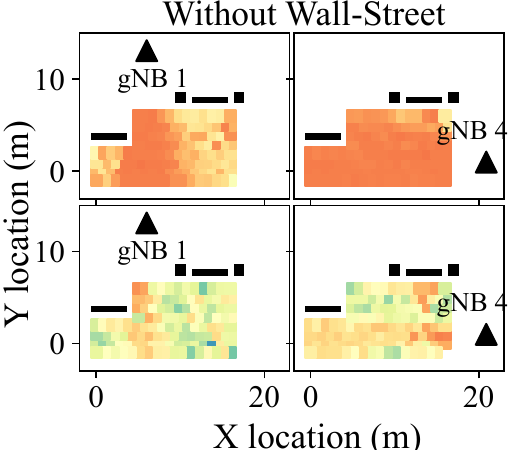}
\caption{\small Baseline single-link}
\label{f:eval:multi_beam:base}
\end{subfigure}
\begin{subfigure}[b]{.52\linewidth}
\includegraphics[width=1\linewidth]{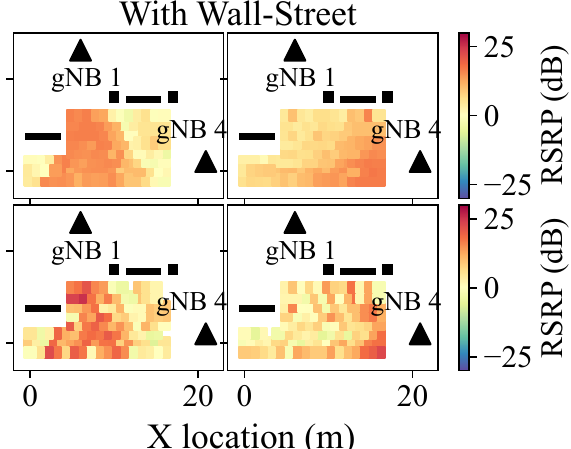}
\caption{\small Wall-Street dual-links}
\label{f:eval:multi_beam:surf}
\end{subfigure}
\caption{\textbf{Indoor multi-link evaluation:} (a) baseline single-link: optimal beam alignment (\textit{upper}), misaligned (\textit{lower}). (b) Wall-Street dual-link: make-before-break with 50/50 power split (\textit{upper}), RSRP inference from reflected signals (\textit{lower}). gNB $\blacktriangle$, blockage $\blacksquare$. }
\label{f:eval:multi_beam}
\end{figure}

\parahead{Multi-link operations.}
\Cref{f:eval:multi_beam} evaluates \shortname{}'s ability to simultaneously control two beams in an indoor environment.
In \cref{f:eval:multi_beam:base}, we show baseline performance without the surface. 
The upper heatmaps display the best-case RSRP from gNB 1 (\textit{left)} or gNB 4 (\textit{right)} with optimal UE beam alignment, 
while the lower heatmaps depict the worst-case scenario where beams are misaligned.
\Cref{f:eval:multi_beam:surf} demonstrates \shortname{}'s dual-link capability. 
The upper heatmaps reveal simultaneous signal reception where \shortname{} splits power equally between gNB1 and gNB4. 
Although individual RSRPs are approximately 3 dB lower than single-beam alignment due to this power division, both links remain active. 
The lower heatmaps demonstrate our RSRP inference method by showing signal strengths calculated from reflected measurements.



\subsection{Throughput and Latency}
\label{s:eval:e2e}
We evaluate throughput and latency performance of \shortname{} delivering bulk TCP data flows. 
To quantify performance, we calculate throughput as the total data bits divided by the duration, measured every 100~ms.
We average the round-trip-time (RTT) of packets over every 100~ms.

\begin{figure}
\begin{subfigure}[b]{1\linewidth}
\includegraphics[width=1\linewidth]{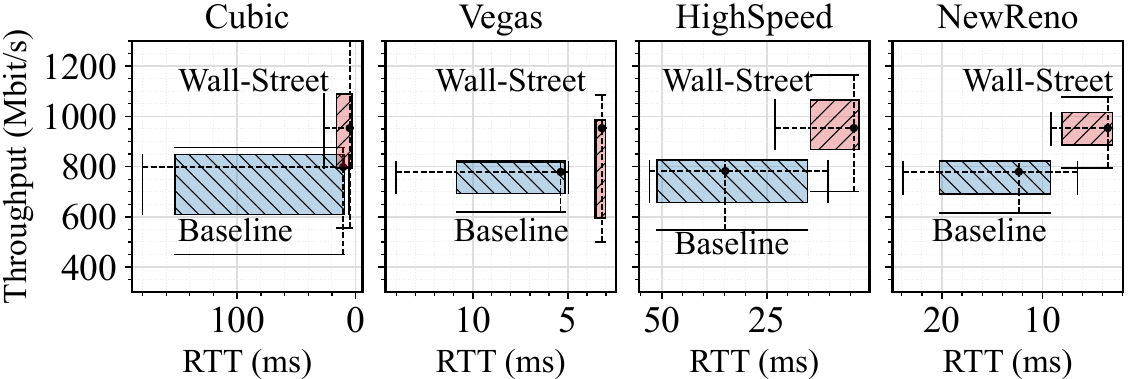}
\caption{Indoor TCP results with one UE.}
\label{f:eval:e2e:indoor}
\end{subfigure}
\begin{subfigure}[b]{1\linewidth}
\includegraphics[width=1\linewidth]{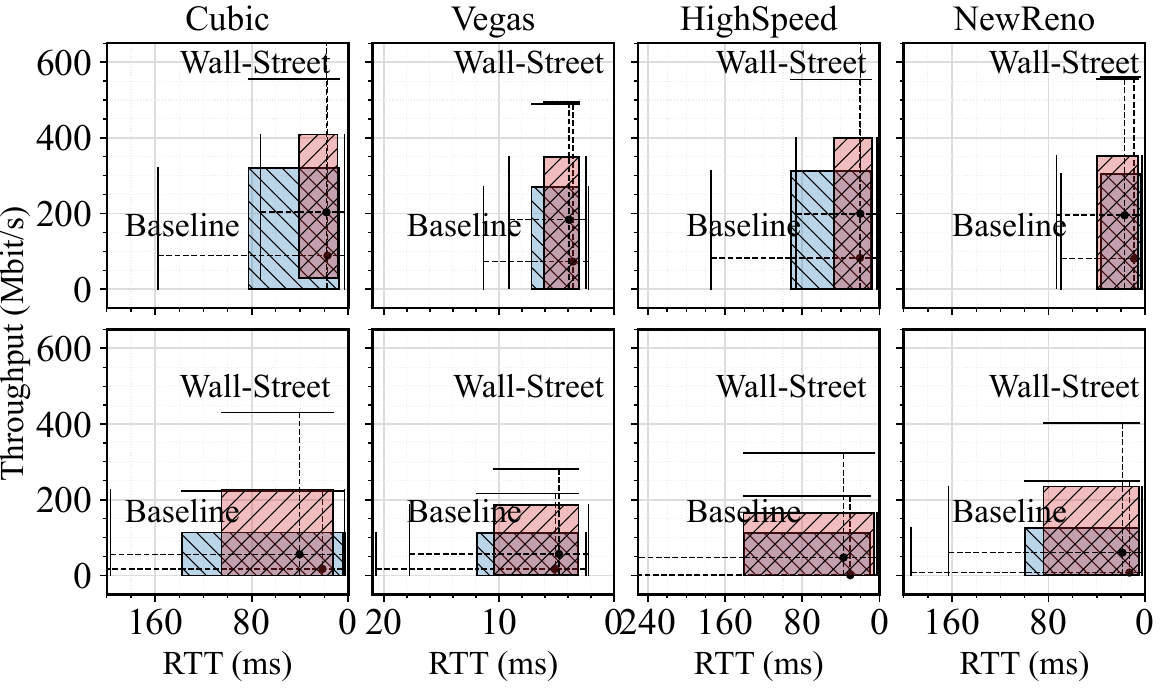}
\caption{Outdoor TCP results with two UEs (\textit{upper:} UE 1; \textit{lower:} UE 2).}
\label{f:eval:e2e:outdoor}
\end{subfigure}
\caption{\textbf{Throughput and round trip time (RTT)} with four congestion control algorithms. The right and lower edge
of the box represents the 10\% percentile, respectively. The left and upper edge give the 90\% percentiles. 
The intersection point 
represents the median value.
}
\label{f:eval:e2e}
\end{figure}

\parahead{TCP performance.}
We evaluate the throughput and Round-Trip Time (RTT) of \shortname{} against the baseline in both indoor and outdoor settings, using four congestion control algorithms: CUBIC, Vegas, HighSpeed, and NewReno.  
For indoor (\cref{f:eval:e2e:indoor}), \shortname{} enhances throughput by at least 20\% and reduces delay by 20-80\%.
Under outdoor multi-user settings (\cref{f:eval:e2e:outdoor}), 
\shortname{} yields more significant improvements. 
Throughput improves by up to 78\% (68\% on average for UE 1 and 60\% on average for UE 2) and delay reduces by up to 34\%  (14\% on average for UE 1 and 5\% on average for UE 2).
Notably, in the baseline setup, UE 2 (located in the cargo area) suffers from lower throughput and higher RTT due to vehicle body blockage

\begin{figure}
\centering
\includegraphics[width=1\linewidth]{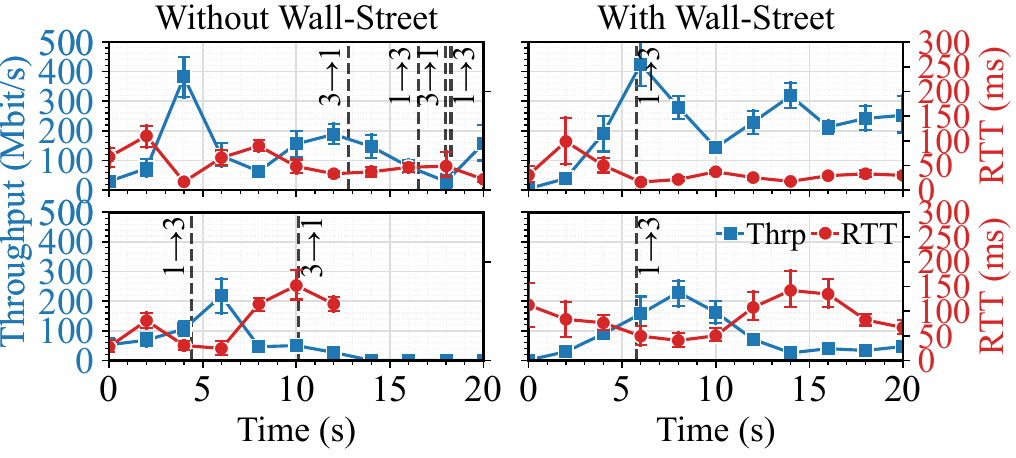}
\caption{\textbf{Throughput and RTT over time} without \shortname{} (\textit{left}) and
with \shortname{} (\textit{right}) for UE 1 (\textit{upper}) and UE 2 (\textit{lower}) as the vehicle moves along the same trajectory. Vertical dashed lines indicate HO triggers.}
\label{f:eval:e2e_realtime}
\end{figure}

\parahead{TCP performance over time.}
\Cref{f:eval:e2e_realtime} divides the 20-second outdoor trajectory into 10 intervals, showing median throughput and delay for UE 1 (\textit{upper}) and UE 2 (\textit{lower}) over time. Vertical dashed lines denote HO triggers.
Wall-Street demonstrates three key advantages over baseline.
First, Wall-Street maintains higher throughput and lower latency throughout the trajectory.
UE 1 achieves higher throughput improvement, and UE 2 maintains continuous connectivity at the 12s mark where the baseline forces a connection termination.
Second, at the 6s mark, Wall-Street executes a collective HO for both UEs. 
In contrast, the baseline triggers separate HOs, causing individual service interruptions and increased signaling overhead.
Lastly, the baseline suffers from ping-pong HOs between gNB 1 and gNB 3. Wall-Street eliminates this instability through robust decision logic.

\begin{figure}[t]
\centering
\includegraphics[width=1\linewidth]{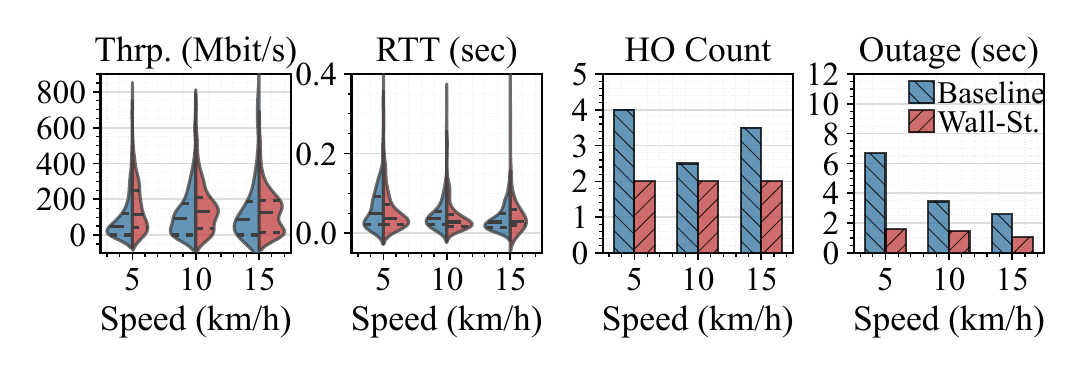}
\caption{\textbf{Impact of speed from 5 to 15 km/h:} throughput, RTT, HO count, and total outage duration (\textit{left} to \textit{right}).}
\label{f:eval:speed}
\end{figure}

\parahead{Impact of vehicle speed.}
\Cref{f:eval:speed} compares \shortnames{} performance against the baseline at driving speeds from 5 to 15 km/h. 
\shortname{} consistently outperforms the baseline across most metrics. Specifically, it boosts user throughput by 70\% at 5 km/h, 27\% at 10 km/h, and 8.4\% at 15 km/h. 
Reliability also improves significantly as \shortname{} reduces HOs by an average of 37\%, lowers latency by 11\%, and cuts connection outages by 64\%.
\shortname{} maintains a consistently low HO count regardless of speed because it makes one conservative decision for all UEs based on multiple measurements, preventing the unnecessary HOs observed in the baseline. 
We note that performance gains diminish at higher speeds due to latency constraints in our current implementation, which limit the speed of HO preparation. 

\begin{figure}[t]
\centering
\includegraphics[width=.92\linewidth]{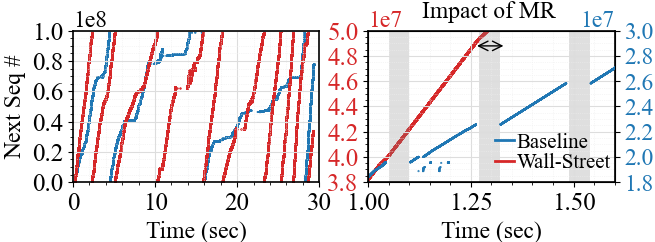}
\caption{\textbf{Measurement report impact}: increase trend of sequence numbers during the entire 30s experiment (left) and handover preparation (right).}
\label{f:eval:hoprep}
\end{figure}

\parahead{Impact of handover preparation.} 
\Cref{f:eval:hoprep} tracks the progression of sequence numbers during the outdoor experiment.
The left subfigure covers the full duration, while the right focuses specifically on the HO preparation phase.
We apply a modulo operation to sequence numbers to visualize trends.
\shortname{} maintains a continuous, steep slope, demonstrating successful packet delivery during HO preparation.
The baseline, however, shows irregular progression and a lower slope because the UE stops data transmission to scan for neighboring gNB scanning.
While SSB detection takes under 20 ms, its effects persist for 50 ms, significantly disrupting baseline packet delivery.

\begin{figure}
\centering
\vspace{-10pt}
\includegraphics[width=.92\linewidth]{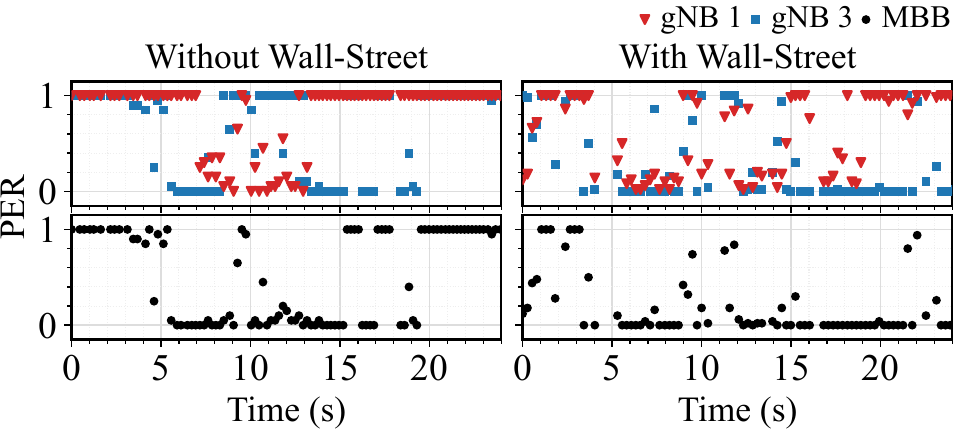}
\caption{\textbf{Real-time make-before-break performance under mobility:} PER of individual links from each gNB (\textit{upper}), and combined PER using duplicate packets from two gNBs (\textit{lower}). Data reflects real-time driving data.}
\label{f:eval:softho}
\end{figure}

\parahead{Impact of make-before-break handover.}
\Cref{f:eval:softho} evaluates Make-Before-Break (MBB) performance via real-time PER measurements over a 25-second drive at 5 km/h. In this setup, gNB 1 and gNB 3 transmit duplicate packet streams for both the Wall-Street and baseline scenarios.
We extract both per-link and combined PER metrics.
The baseline setup is inherently limited by the UE's single-beam constraint. 
Even though both gNBs are transmitting, the UE can only align its beam with one (here, gNB 3). Consequently, packets from gNB 1 rarely reach the UE, effectively negating the benefits of MBB.
In contrast, \shortname{} simultaneously steer both gNB signals to the UE.  
While individual links (red/blue) operate at moderate PER due to power splitting, their combined reception (black) achieves consistently low PER.

\begin{figure}[t]
\centering
\includegraphics[width=.75\linewidth]{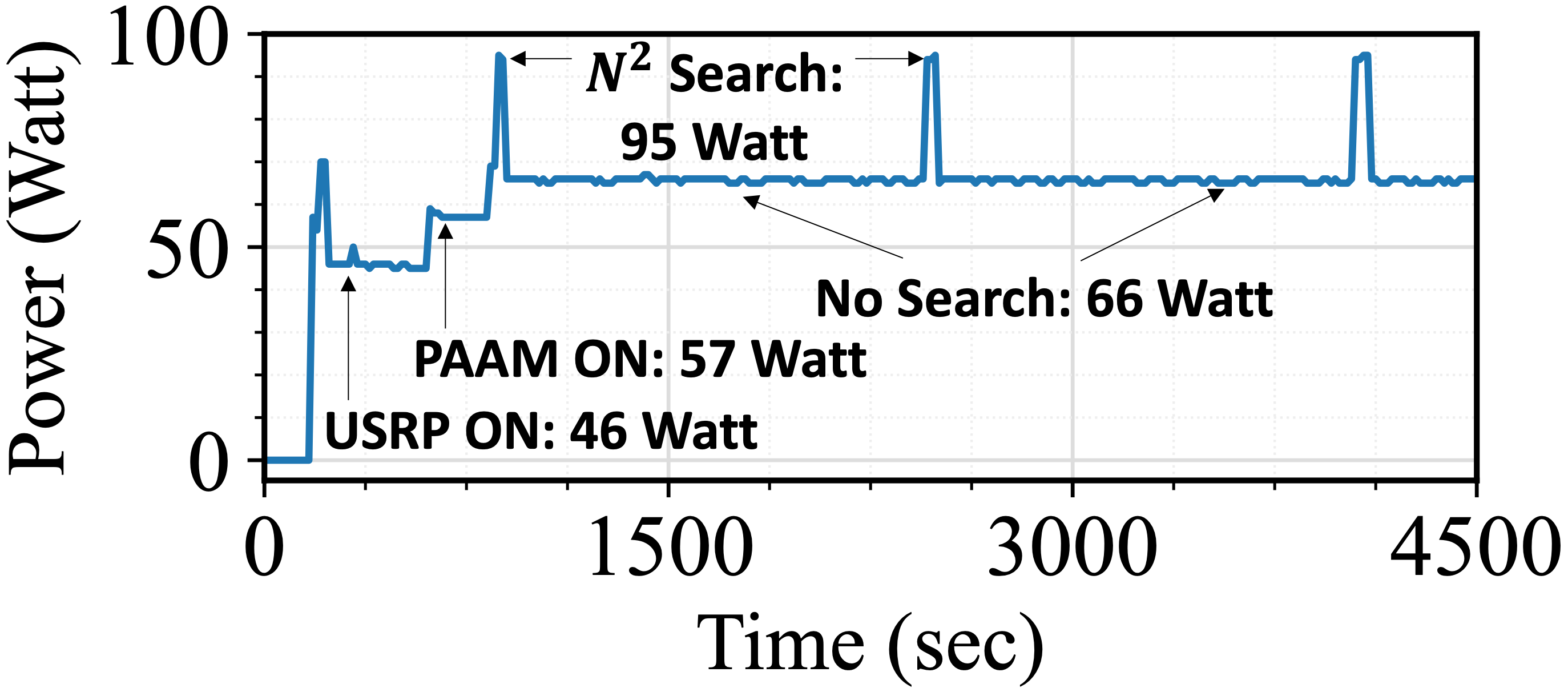}
\caption{\textbf{UE power measurement}: power increases from 66 W to 95 W during the search process.}
\label{f:eval:power}
\end{figure}

\parahead{Power Consumption.}
Our measured power consumption ranges from 0.48 W to 0.64 W. This is largely due to the 16 V supply required for the two COTS DAC controllers~\cite{noauthor_analog_nodate} to drive 76 boards (drawing from 0.03 to 0.04 A in total). The surface itself consumes only $\approx160$ $\micro$W. 
We can significantly lower power consumption by using custom low-power DACs or redesigning the surface to achieve full 2$\pi$ phase shifts with 0 - 8 V, which would reduce the power consumption by half.
We profiled the power consumption of COSMOS UE nodes using an external measurement tool. 
To accommodate the power measurement tool's 15-sec sampling resolution, we time-scaled the alignment operations from ms to s.
\cref{f:eval:power} shows that while the UE consumes 66 W without scanning, this surges to 95 W during the exhaustive beam search. \shortname{} effectively eliminates this peak by bypassing scanning.

\section{Limitations and Discussions}
\label{s:discussion}
To integrate our system with cellular infrastructure, we make several non-standard compliant assumptions.
In this section, we outline each assumption and discuss potential solutions.

\parahead{gNB-side Measurement.}
Our system assumes the serving gNB can measure reflected signals from a neighbor gNB while serving the UE. To adopt our system under TDD deployment, we can synchronize gNBs using inter-gNB coordination. 
We can use the standard Xn interface, which allows gNBs to exchange configuration data, including TDD slot formats and SSB periodicity. 
The serving gNB can acquire the neighbor gNB's broadcast schedule relative to its own internal clock and align an uplink slot to overlap with the neighbor’s SSB transmission. During this slot, the serving gNB is in receive mode and measures the Wall-Street reflection.
We can also implement our system into an O-RAN architecture, offloading this coordination logic to the O-RAN Near-Real-Time RAN Intelligent Controller (Near-RT RIC).


\parahead{Control Channel.}
Our system uses an out-of-band sub-6 GHz link to integrate with the COSMOS roadside testbed, as the testbed requires all components to be connected through the same local network. 
Here, \shortname{} receives a few control bits to index its pre-computed voltage codebook. 
In deployment, this control information can move in-band. The gNB can embed low-rate control data directly into the mmWave SSB it already transmits, using simple modulation schemes such as Frequency-Shift Keying (FSK) on a subcarrier. The surface would then need only a low-power detector synchronized to the SSB periodicity to decode these commands, eliminating the need for a sub-6 modem entirely.

\parahead{Reducing Alignment Overhead.}
Initial beam alignment requires high overhead across the gNB, surface, and UE.
We can reduce this complexity in two ways. 
First, we leverage vehicle GPS for coarse location estimation, allowing the gNB to narrow its transmit beam search sector. 
We note that we may use GPS for accelerating beam search, but not for handover decisions as it cannot capture environmental blockages or vehicle body effects that determine actual signal quality. 
Second, the surface can operate as a reflective probe. 
It can reflect and steer the gNB's signals while the gNB listens for the strongest return, taking $O(N^2)$ to determine their mutual orientation. 
The surface can then reflect and steer UE signals toward the UE with another $O(N^2)$ search. 


\parahead{Deployment Feasibility.}
We propose mounting the surface on the vehicle’s interior sunroof or glass ceiling. This placement aligns with the arrival angle of downlink signals while protecting the hardware from rain and snow.
Our link-budget analysis (Appendix~\ref{appendix:linkbudget}) confirms that a compact $10\times20$ cm surface provides sufficient signal strength for both user connectivity and cell search under real-world settings.

\parahead{Scalability.} We address scalability across two dimensions:
\begin{itemize}
    \item \textbf{Interference:} As the number of vehicles increases, surface reflections could interfere with nearby networks or outdoor users. Since the metamaterial surface use tight element spacing, precise phase control could further nullify interference. Also, because gNBs transmit downwards onto the roof-mounted surface, reflections are directed upward, minimizing leakage to ground-level outdoor users.
    \item \textbf{UE contention:} Wall-Street uses time-division to serve multiple users, which causes contention as the number of UEs increases. We may partition the surface into multiple sub-areas to serve multiple users simultaneously when not performing handovers. A MAC-layer scheduler can dynamically manage these resources, allocating larger surface areas to users with higher throughput demands.
\end{itemize}

\parahead{Complex Mobility.}
Our outdoor experiments were constrained by the testbed’s fixed topology and control network access. Real-world driving involves unpredictable turns and varying speeds. At a distance of 150 meters with a standard 15\textdegree{} beamwidth, the mmWave beam footprint on the ground is approximately 40 m wide. This is sufficient to cover a bus even during quick turns. 
We may also use GPS data for beam tracking to add robustness against rapid speed changes.

\section{Conclusion}
\label{s:concl}
This paper introduces Wall-Street that efficiently steers outdoor mmWave signals into vehicles, simultaneously exchange data and and scan neighboring cells, and realizes make-before-break HOs. We integrated the surface into the COSMOS testbed and conducted driving experiments with surface-mounted SUV and multiple gNB/UE nodes. Our trace-driven ns-3 simulations show the TCP improvement by up to 78\% and latency reduction by up to 34\%. 
\color{black}

\section*{Acknowledgement}
This material is based upon work supported by the National Science Foundation under Grant Nos. CNS-2148271, CNS-2433915, CNS-2450567 and is supported in part by funds from federal agency and industry partners as specified in the Resilient \& Intelligent NextG Systems (RINGS) program. We gratefully acknowledge the support of a grant from the Princeton University School of Engineering and Applied Science Innovation Fund.


\appendix


\begin{figure}
\begin{subfigure}{.8\linewidth}
\includegraphics[width=1\linewidth]{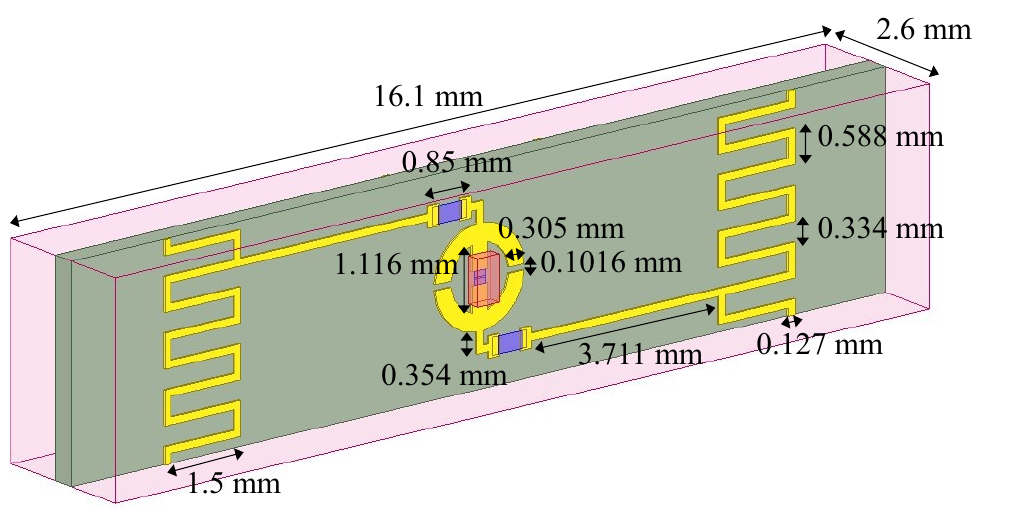}
\caption{Electric meta-atom.}
\end{subfigure}
\begin{subfigure}{.8\linewidth}
\includegraphics[width=1\linewidth]{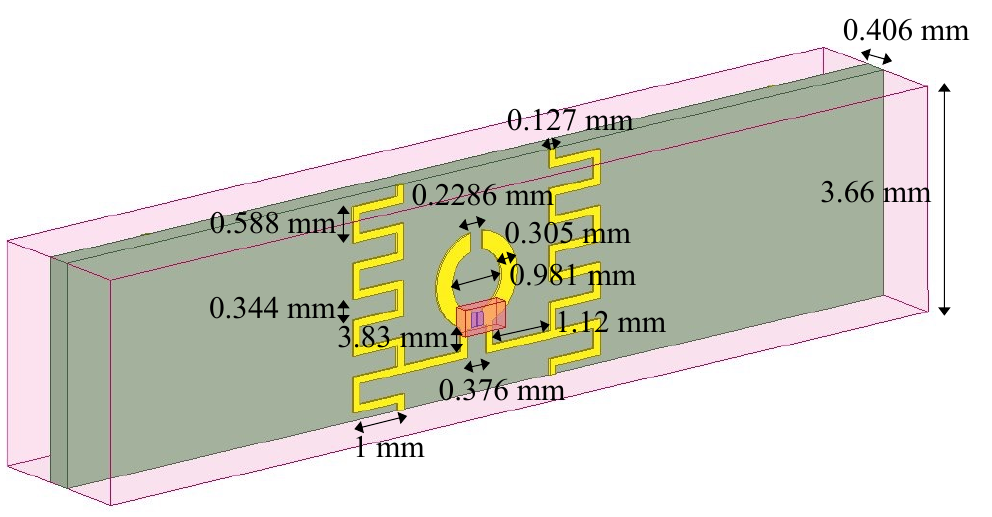}
\caption{Magnetic meta-atom.}
\end{subfigure}
\caption{\systemnames{} unit cell design parameters.}
\label{f:surface_param}
\end{figure} 

\section{Surface Design Parameters}
\label{appendix:surface_param}
In this section, we provide design parameters of Wall-Street's electric and magnetic meta-atom, as illustrated in \cref{f:surface_param}. 

\section{Link Budget Calculation}
\label{appendix:linkbudget}

To demonstrate the deployment feasibility, we perform a link-budget analysis using a realistic in-vehicle deployment scenario. 
First, we calculate the downlink SNR at the in-vehicle UE under worst-case conditions:
\begin{equation}
\begin{aligned} 
SNR_{UE}~(dB) = P_{gNB} - L_{gNB} - L_{window} + G_{RIS,Rx} + \\ G_{RIS,Tx} - L_{UE} + G_{UE} - P_{nf}
\end{aligned} 
\end{equation}
We consider a terrestrial base station with an EIRP ($P_{gNB_s}$) of 60 dBm~\cite{qualcomm} and a noise floor ($P_{nf}$) of -89 dBm. The worst-case distance from the base station to the vehicle is set to 0.15 km~\cite{hassan_vivisecting_2022}, corresponding to a free-space path loss ($L_{gNB}$) of 103 dB. 
Inside the mid-sized vehicle like bus, we assume the RIS is mounted on the mid-roof at the center, with a maximum distance to the user of 4 m, resulting in an internal path loss ($L_{UE}$) of 72 dB. The roof window introduces an additional penetration loss ($L_{window}$) of 3 dB~\cite{glass}.
The RIS contributes to the link budget in two stages. First, the aperture gain ($G_{RIS,Rx}$) accounts for the physical size of the RIS capturing energy. 
Second, the beamforming gain ($G_{RIS,Tx}$) accounts for the RIS phasing the signal to focus it on the user~\cite{9998527, basar2019wireless, cho_mmwall_2023, 8811733}\footnote{\textcolor{black}{The SNR gain from an N-element RIS is proportional to $N^2$~\cite{9998527, basar2019wireless, cho_mmwall_2023, 8811733}.}}\footnote{\textcolor{black}{In Sec.~\ref{s:design:ho:decision}, we collectively account the surface gains as a total gain $G_{w}$.}}.
For our 10$\times$20 cm surface, the gain varies between 24 to 29 dBi depending on the steering angle, and we adopt the conservative lower bound of 24 dBi.
Assuming a typical mmWave smartphone receiver gain ($G_{UE}$) of 8 dBi~\cite{koul2020compact}, the resulting SNR is approx. 27 dB, which is a sufficient link margin for high-throughput connectivity.

We also evaluate the feasibility of gNB-to-gNB measurement, where the serving gNB measures a signal from the neighbor gNB reflected by the in-vehicle RIS. 
We calculate the reflective SNR measured at the serving gNB under worst-case conditions as:
\begin{equation} 
\label{eq:snr_gnb} 
\begin{aligned} 
SNR_{gNB}~(dB)= P_{gNB_n} - L_{gNB_n} - L_{window} + G_{RIS,Rx} + \\ G_{RIS,Tx} - L_{window} - L_{gNB_s} + G_{gNB_s} - P_{nf} 
\end{aligned} 
\end{equation}
where $P_{gNB_n}$ is a neighbor gNB's EIRP of 60 dBm, and $L_{gNB_n}$ and $L_{gNB_s}$ are the worst-case distance between the vehicle to the neighbor gNB and the serving gNB, respectively, which is 103 dB.
In this scenario, the signal traverses the car window twice (entry and exit), doubling the penetration loss ($L_{window}$). Assuming a serving gNB receive gain ($G_{gNB_s}$) of 29.5 dB~\cite{arunruangsirilert2025performance}, we achieve a minimum SNR of approx. 14.5 dB with \shortname{}. 
Our analysis confirms that the 10$\times$20 cm surface size is sufficient to support both user connectivity and cell measurement in realistic scenarios.

\clearpage
\let\oldbibliography\thebibliography
\renewcommand{\thebibliography}[1]{%
  \oldbibliography{#1}%
  \setlength{\parskip}{0pt}%
  \setlength{\itemsep}{0pt}%
}
\bibliographystyle{ACM-Reference-Format} 
\bibliography{zotero-paws}

\end{document}